\documentclass[camera,letterpaper,nomarginnotes,nonarrowgutter]{jpaper}
\usepackage[compress,sort]{cite}

\usepackage{amsmath,amssymb,amsfonts}
\usepackage{graphicx}
\usepackage{textcomp}
\usepackage{xcolor}
\usepackage{xspace}
\usepackage{fancyhdr}
\usepackage[normalem]{ulem}
\usepackage{booktabs}
\usepackage{multirow}
\usepackage{setspace}
\usepackage{titlesec}
\usepackage{tikz}
\usepackage{enumitem}
\usepackage{pifont}
\usepackage{mathtools}
\usepackage{dblfloatfix} 
\usepackage[bottom]{footmisc}
\usepackage{balance}
\usepackage{xargs}  
\usepackage{setspace}
\usepackage{datetime}

\usepackage{mathptmx} %
\usepackage[keeplastbox]{flushend}
\usepackage[bookmarks=true,breaklinks=true,letterpaper=true,colorlinks,citecolor=blue,linkcolor=blue,urlcolor=blue]{hyperref}

\newcommand\js[1]{{\color{black}{}{#1}}}

\definecolor{brightlavender}{rgb}{0.75, 0.58, 0.89}

\newcommand\fixme[1]{{\color{black}{}{#1}}}

\definecolor{bondiblue}{rgb}{0.0, 0.58, 0.71}
\definecolor{cardinal}{rgb}{0.77, 0.12, 0.23}
\definecolor{carminered}{rgb}{1.0, 0.0, 0.22}
\definecolor{darkorchid}{rgb}{0.6, 0.2, 0.8}

\newcommand\fixok[1]{{\color{black}{}{#1}}}
\newcommand\passthree[1]{{\color{black}{}{#1}}}
\newcommand\passfour[1]{{\color{black}{}{#1}}}
\newcommand\passfive[1]{{\color{black}{}{#1}}}
\newcommand\passsix[1]{{\color{black}{}{#1}}}

\newcommand\prop{Flash-Cosmos\xspace}
\newcommand\esp{ESP\xspace}
\newcommand\mws{MWS\xspace}
\newcommand\espc{\texttt{ESP}\xspace}
\newcommand\mwsc{\texttt{MWS}\xspace}
\newcommand\bbo{bulk bitwise operation\xspace}
\newcommand\bbos{bulk bitwise operations\xspace}

\newcommand\cbbos{Bulk bitwise operations\xspace}
\newcommand\inum[1]{(\textit{#1})\xspace}
\newcommand{\sect}[1]{{Section~#1}\xspace} %
\newcommand{\head}[1]{{\noindent\textbf{#1.}\xspace}} %
\newcommand{\figs}[1]{{Figures~#1}\xspace} %
\newcommand{\fig}[1]{{Figure~#1}\xspace} %
\newcommand{\tab}[1]{{Table~#1}\xspace} %
\newcommand{\subs}[2]{#1$_{\text{#2}}$\xspace}%
\DeclareRobustCommand\bcirc[1]{\tikz[baseline=(char.base)]{
           \node[shape=circle,draw,inner sep=0pt,fill=black, text=white] (char) {#1};}}
\DeclareRobustCommand\wcirc[1]{\tikz[baseline=(char.base)]{
           \node[shape=circle,draw,inner sep=1pt,fill=white, text=black] (char) {#1};}}
\makeatletter
\newcommand*\bigcdot{\mathpalette\bigcdot@{.5}}
\newcommand*\bigcdot@[2]{\mathbin{\vcenter{\hbox{\scalebox{#2}{$\m@th#1\bullet$}}}}}
\makeatother

\newcommandx{\changev}[2][1=]{\todo[linecolor=blue,backgroundcolor=blue!25,bordercolor=blue,#1,size=\scriptsize]{#2}}

\newcommand{\vth}{V$_{\text{TH}}$\xspace}
\newcommand{\vthn}{V$_{\text{TH}}$}
\newcommand{\vpre}{\subs{V}{PRE}}
\newcommand{\vref}{V$_{\text{REF}}$\xspace}
\newcommand{\vrefn}[1]{V$_{\text{REF#1}}$}
\newcommand{\vpass}{V$_{\text{PASS}}$\xspace}

\newcommand{\vprog}[1]{V$_{\text{PGM}#1}$}
\newcommand{\vispp}{$\Delta$V$_{\text{ISPP}}$}
\newcommand{\vtarget}{V$_{\text{TGT}}$}
\newcommand{\cso}{\subs{C}{SO}}
\newcommand{\tbers}{\texttt{tBERS}\xspace}
\newcommand{\tr}{\texttt{tR}\xspace}
\newcommand{\tprog}{\texttt{tPROG}\xspace}
\newcommand{\tdma}{\texttt{tDMA}\xspace}
\newcommand{\textr}{\texttt{tEXT}\xspace}
\newcommand{\tmws}{\texttt{tMWS}\xspace}
\newcommand{\tesp}{\texttt{tESP}\xspace}

\newcommand{\bnot}{\texttt{NOT}\xspace}
\newcommand{\band}{\texttt{AND}\xspace}
\newcommand{\bor}{\texttt{OR}\xspace}
\newcommand{\bxor}{\texttt{XOR}\xspace}
\newcommand{\bnand}{\texttt{NAND}\xspace}
\newcommand{\bnor}{\texttt{NOR}\xspace}
\newcommand{\bxnor}{\texttt{XNOR}\xspace}

\newcommand{\bxorxy}[2]{\texttt{#1~XOR~#2}\xspace}

\newcommand{\bxnorxy}[2]{\texttt{#1~XNOR~#2}\xspace}

\newcommand{\usec}{\textmu{}s\xspace} 
\newcommand{\pbit}{ParaBit\xspace}
\newcommand{\ol}[1]{$\overline{#1}$}

\newcommand{\out}[1]{OUT$_\text{#1}$}
\newcommand{\nout}[1]{\ol{\text{OUT}_\text{#1}}}
\newcommand{\degreec}[1]{#1$^\circ$C\xspace}
\newcommand{\osp}{\texttt{OSP}\xspace}
\newcommand{\isp}{\texttt{ISP}\xspace}
\newcommand{\pb}{\texttt{PB}\xspace}
\newcommand{\fc}{\texttt{FC}\xspace}
\newcommand{\bmi}{\textsf{BMI}\xspace}
\newcommand{\ims}{\textsf{IMS}\xspace}

\newcommand{\kcsl}{\textsf{KCS}\xspace}
\def\BibTeX{{\rm B\kern-.05em{\sc i\kern-.025em b}\kern-.08em
    T\kern-.1667em\lower.7ex\hbox{E}\kern-.125emX}}

\fancypagestyle{firstpage}
{
}

\title{\huge \prop: In-Flash \fixok{Bulk} Bitwise Operations Using \\ Inherent Computation Capability of NAND Flash Memory}

\newcommand{\affilETH}[0]{\textsuperscript{\S}}
\newcommand{\affilPostec}[0]{\textsuperscript{$\nabla$}}
\newcommand{\affilCNRS}[0]{\textsuperscript{$\dagger$}}
\newcommand{\affilKyung}[0]{\textsuperscript{$\ddagger$}}

\author{
{Jisung~Park\affilETH\affilPostec}~~~%
{Roknoddin~Azizi\affilETH}~~~%
{Geraldo F. Oliveira\affilETH}~~~%
{Mohammad~Sadrosadati\affilETH}\\
{Rakesh~Nadig\affilETH}~~~%
{David~Novo\affilCNRS}~~~%
{Juan~Gómez-Luna\affilETH}~~~%
{Myungsuk~Kim\affilKyung}~~~%
{Onur Mutlu\affilETH}\\\\%
\emph{{ 
\affilETH ETH Z{\"u}rich~~~~~ \affilPostec POSTECH ~~~~~ \affilCNRS LIRMM, Univ. Montpellier, CNRS ~~~~~
\affilKyung Kyungpook National University}%
}%
}

\begin{document}
\maketitle
\thispagestyle{firstpage}

\begin{abstract}
\cbbos, i.e., bitwise operations on large bit vectors, are prevalent in a wide range of important application domains\fixok{,} including databases\fixok{,} graph processing, genome analysis\fixok{, cryptography,} and hyper-dimensional computing.
In conventional systems, the performance and energy efficiency of bulk bitwise operations are bottlenecked by data movement between the compute units (e.g., CPUs and GPUs) and the memory hierarchy.
In-flash processing (i.e., processing data inside NAND flash chips) has a high potential to accelerate bulk bitwise operations by fundamentally \fixok{reducing} data movement through the entire memory hierarchy, especially when the processed data does not fit into main memory.

\fixok{W}e identify two key limitations of the state-of-the-art in-flash processing technique for bulk bitwise operations;
\inum{i}~it \fixok{falls short of maximally exploiting} the bit-level parallelism of bulk bitwise operations \fixok{that could be \fixok{enabled} by} leveraging the unique cell-array architecture and operating principles of NAND flash memory;
\inum{ii}~\fixok{it is \fixok{unreliable because it is} not designed to take into account the
highly %
error-prone nature of NAND flash memory.} 

We propose \prop (\underline{\fixok{Flash}} \underline{C}omputation with \underline{O}ne-\underline{S}hot \underline{M}ulti-\underline{O}perand \underline{S}ensing), a new in-flash processing technique that significantly increases the performance and energy efficiency of bulk bitwise operations while providing high reliability.
\prop introduces two key mechanisms that can be easily supported in \fixok{modern} NAND flash chips: \inum{i}~\underline{M}ulti-\underline{W}ordline \underline{S}ensing (\mws), which enables bulk bitwise operations on a large number of operands \fixok{(tens of operands)%
} with a \emph{single} sensing operation, and \inum{ii}~\underline{E}nhanced \underline{S}LC-mode \underline{P}rogramming (\esp), which 
\fixok{enables reliable computation inside NAND flash memory.} 
We demonstrate the feasibility \fixok{of performing \bbos with high reliability in} \fixok{\prop by testing} 160 real 3D NAND flash chips.
Our evaluation shows that \prop improves average performance and energy efficiency by 3.5$\times$\fixok{/32$\times$} and 3.\fixok{3}$\times$\fixok{/95$\times$}, respectively, over the state-of-the-art in-flash\fixok{/outside-storage processing  techniques} across three real-world applications.
\end{abstract}

\section{Introduction}\label{sec:introduction}

\fixok{M}any data-intensive applications rely on \emph{\bbos}, i.e., bitwise operations on large bit vectors\fixok{. As such,} it is important for modern computing systems to support high-performance and energy-efficient \bbos. 
In databases \fixok{and web search}, prior works (e.g.,~\cite{chan-signmod-1998, oneil-ideas-2007, li-vldb-2014, li-sigmod-2013, goodwin-SIGIR-2017, seshadri-micro-2013, seshadri-micro-2017, seshadri-ieeecal-2015, hajinazar-asplos-2021}) propose various techniques that 
heavily use \bbos to accelerate queries.
\cbbos are also prevalent in \fixok{various} other important application domains, including databases and web search~\cite{chan-signmod-1998, oneil-ideas-2007, li-vldb-2014, li-sigmod-2013, goodwin-SIGIR-2017, seshadri-micro-2013, seshadri-micro-2017, seshadri-ieeecal-2015, hajinazar-asplos-2021, FastBitA9, wu-icde-1998, guz-ndp-2014,redis-bitmaps},
\passthree{data analytics~\cite{perach-arxiv-2022, seshadri-micro-2017, jun-isca-2015, torabzadehkashi-pdp-2019, lee-ieeecal-2020},}
\fixok{graph processing~\cite{beamer-SC-2012, besta-micro-2021, li-dac-2016, gao-micro-2021, hajinazar-asplos-2021}, genome analysis~\cite{alser-bioinformatics-2017, loving-bioinformatics-2014, xin-bioinformatics-2015, cali-micro-2020, kim-genomics-2018, lander-nature-2001, altschul-jmb-1990, myers-jacm-1999}, cryptography~\cite{han-spie-1999, tuyls-springer-2005, manavski-spcom-2007}, set operations~\cite{besta-micro-2021, seshadri-micro-2017},} and hyper-dimensional computing~\cite{kanerva-1992, kanerva-congcomp-2009, karunaratne-nature-2020, imani-hpca-2017}.

In conventional systems, the performance and energy efficiency of \bbos are bottlenecked by \emph{data movement} between the compute units (e.g., CPUs or GPUs) and the memory hierarchy~\cite{aga-hpca-2017, seshadri-ieeecal-2015, seshadri-micro-2017, gao-micro-2021, li-dac-2016, seshadri-arxiv-2019,li-micro-2017}.
To perform a \fixok{bulk} bitwise operation, \fixok{a} conventional system must first move every operand to the compute unit and \fixok{eventually} write the results \fixok{back into} the memory hierarchy.
Due to the simple nature of bitwise operations, such data movement dominates the execution time and energy consumption in \bbos.

Processing data \emph{inside} NAND flash chips, i.e., \emph{in-flash processing (IFP)}, can fundamentally \fixok{reduce} the data movement \fixok{that bottlenecks the execution of} bulk bitwise operations.
IFP is an instance of \emph{near-data processing (\fixok{NDP})}, a computing paradigm that moves computation closer to where the data resides (e.g.,~\cite{aga-hpca-2017, fujiki-isca-2019, eckert-isca-2018, ramanathan-micro-2020, seshadri-ieeecal-2015, seshadri-micro-2017, augusta-sigmod-2015, ahn-isca-2015, ahn-isca-2015-2, drumond-isca-2017, farmahini-hpca-2015, gao-hpca-2016, hajinazar-asplos-2021, li-dac-2016, jun-isca-2018, mailthody-micro-2019, seshadri-osdi-2014, gu-isca-2016, gao-micro-2021, balasubramonian-micro-2014, mutlu-emergingcomputing-2021}).
When processing large amount\fixok{s} of data that do not fit in main memory,
IFP significantly reduces data movement \fixok{across} the entire memory hierarchy by performing computation within the underlying storage media (i.e., NAND flash chips) and transferring only the result \fixok{(when needed, to main memory and CPUs/GPUs)}. 
\fixok{A}s we discuss in \sect{\ref{sec:motiv}}, IFP can significantly outperform in-storage processing (ISP) approaches that leverage hardware accelerators inside the NAND flash-based solid-state drive (SSD) (e.g.,~\cite{jun-isca-2018, mailthody-micro-2019, seshadri-osdi-2014, gu-isca-2016,mansouri-asplos-2022,kang-msst-2013}), by \fixok{reducing} data movement \fixok{to/}from NAND flash chips.

To our knowledge, only \fixok{one} recent work, \pbit~\cite{gao-micro-2021}, proposes an in-flash processing technique for \bbos.\footnote{
There are many prior works \fixme{(e.g.,~\cite{choi-iscas-2020, han-ieeetcas-2019, shim-ieeejetcas-2022, bayat-ieeetnnls-2017, tseng-iedm-2020, wang-ieeetvlsi-2018, lue-iedm-2019})} that leverage analog current sensing to perform accumulative computation (e.g., multiply-accumulate operation) inside NAND flash chips.
However, \fixok{these} proposals \fixok{1)} use NAND flash memory \fixok{solely} as an accelerator but \emph{not} as a storage medium, \fixok{and 2)} require significant changes (e.g., adding \fixok{a} precise analog-to-digital converter to each bitline) to commodity NAND flash chips\fixok{, which increases cost}. 
See \sect{\ref{sec:related}} for more detail.}  
\fixok{W}e identify \fixok{that \pbit has} two \fixok{major} limitations.
First, \pbit \fixok{ falls greatly short of exploiting the full} potential of NAND flash memory to significantly improve the performance and energy efficiency of \bbos.
To perform \fixok{\bbos} for more than two operands (e.g., \texttt{A}~$\bigcdot$~\texttt{B}~$\bigcdot$~\texttt{C}), 
which frequently happens in many data-intensive applications such as data analytics\fixok{~\cite{perach-arxiv-2022, seshadri-micro-2017, jun-isca-2015, torabzadehkashi-pdp-2019, lee-ieeecal-2020}, databases~\cite{FastBitA9, wu-icde-1998, guz-ndp-2014,redis-bitmaps}} and graph processing~\cite{li-dac-2016, gao-micro-2021,besta-micro-2021, hajinazar-asplos-2021}, 
\pbit must \emph{serially} perform multiple two-operand bitwise operations (e.g., (\texttt{A}~$\bigcdot$~\texttt{B})~$\bigcdot$~\texttt{C}).
Doing so requires multiple \fixok{long-latency} sensing operations \fixok{in series}, which become a new performance \fixok{and energy efficiency} bottleneck.
\fixok{In this work}, we \fixok{observe} that NAND flash memory has \emph{inherent} capability to perform bitwise operations on a large number \fixok{(e.g., tens)} of operands \emph{at once} \fixok{(i.e., with a single sensing operation)} due to its unique cell-array structures that are similar to digital logic circuits for \bnand and \bnor gates.   

Second, \pbit is applicable only to \emph{highly error-tolerant} applications \fixok{because it is \passthree{\emph{not}} designed to take into account} the \fixok{highly} error-prone nature of NAND flash memory.
To ensure data reliability, modern NAND flash-based SSDs commonly use \inum{i} error-correcting codes (ECC) and \inum{ii} data randomization~\cite{park-asplos-2021, cai-insidessd-2018, cha-etrij-2013, cai-procieee-2017}.
Unfortunately, \pbit cannot leverage any of the widely-used ECC and data-randomization techniques, as it performs bitwise operations \emph{while} sensing the cells that store the data. Performing bitwise \band and \bor operations on ECC-encoded or randomized data \fixok{using \pbit} can lead to incorrect results \fixok{during} ECC decoding and\fixok{/or} de-randomization.
Although storing a \fixok{smaller} number of bits in a cell would reduce the raw bit error rate (RBER) of NAND flash memory, our characterization using \fixok{160} real 3D NAND flash chips shows that even storing a single bit per cell \emph{cannot} provide sufficiently-low RBER for \pbit to be adopted \fixok{across} a wide range of applications.

\textbf{Our goal} is to improve both the performance and energy efficiency \fixok{of in-flash} \bbos while \fixok{ensuring} \fixok{high reliability (i.e., zero bit errors)} \fixok{in computation results.}
To this end, we propose \emph{\prop (\fixok{\underline{Flash}} \underline{C}omputation with \underline{O}ne-\underline{S}hot \underline{M}ulti-\underline{O}perand \underline{S}ensing)}, a novel in-flash processing technique for \bbos \fixok{ that} achieves our goal by exploiting\linebreak \textbf{two key ideas:} \inum{i}~\emph{Multi-Word\-line Sensing (\mws)}, which enables in-flash \bbos on multiple \fixok{(e.g., tens)} operands with a \emph{single} sensing operation, and \inum{ii}~\emph{Enhance\fixok{d} SLC-mode Programming (\esp)}, which effectively achieves \emph{zero} bit errors \fixok{in the results of} \fixok{in-flash \bbos}.

\mws leverages the two fundamental cell-array structures of NAND flash memory to perform in-flash \bbos on a large number of operands with a \emph{single} sensing operation\fixok{:}
\inum{i}~a number of flash cells (e.g., 24--176 cells) are serially connected \fixok{to form} a NAND string (similar to digital \bnand logic); 
\inum{ii}~thousands of NAND strings are connected to the same bitline (similar to digital \bnor logic).
Under these cell-array structures, simultaneously sensing \emph{multiple} wordlines\footnote{
NAND flash memory concurrently reads a large number of ($> 10^5$) cells whose control gates are connected to the same wordline. See \sect{\ref{ssec:bg_basics}} for more \fixok{background on NAND flash \passthree{memory} operation}.} 
automatically results in \inum{i}~bitwise \band of \emph{all} the sensed wordlines if they are in the same NAND string or \inum{ii}~bitwise \bor of \emph{all} the wordlines if they are in different NAND strings.

\esp effectively avoids raw bit errors in stored data via more precise programming-voltage control.
A flash cell stores bit data \fixok{as a function of the level of} its threshold-voltage (\vth). Reading a cell incurs an error if the cell's \vth level moves to another \vth range that \fixok{corresponds to} a different bit value \fixok{than} \fixok{the stored value,} due to various reasons~\cite{cai-procieee-2017}, such as program interference~\cite{cai-iccd-2013, cai-hpca-2017, park-dac-2016}, \fixok{data} retention loss~\cite{cai-hpca-2015, cai-iccd-2012, cai-inteltechj-2013, luo-sigmetrics-2018}, \fixok{read disturbance~\cite{cai-dsn-2015, ha-ieeetcad-2015}, and cell-to-cell interference~\cite{cai-hpca-2017}}.
\esp maximizes the margin between different \vth ranges \fixok{by carefully} leveraging two existing approaches.
First, to store data for in-flash processing, it uses the single-level cell (SLC)-mode programming scheme~\cite{lee-atc-2009, kim-asplos-2020}. %
Doing so \fixok{guarantees} a large \vth margin by forming only two \vth ranges (for encoding `\texttt{1}' and `\texttt{0}') within the fixed \vth window. 
Second, \esp enhances the SLC-mode programming scheme by using \inum{i}~a higher programming voltage to increase the distance between the two \vth ranges and \inum{ii}~more programming steps to narrow the high \vth range.  
While many prior works also leverage precise programming to enhance the reliability of NAND flash memory~\cite{jeong-fast-2014, shim-micro-2019, kim-glsvlsi-2018, feng-iccd-2017, Wang-ieeetcom-2016, dong-ieeetcas-2010}, we aim to achieve \emph{zero} bit errors \fixok{in computation results} and demonstrate \fixok{that doing so is possible} in modern NAND flash memory by combining the two approaches \fixok{that comprise ESP}.

\fixok{In this paper, we enhance our basic \mws mechanism in two ways to make it more general purpose.} %
First, \fixok{we} support bitwise \bnand/\bnor/\bxor/\bxnor by using \mws along with \inum{i}~the \emph{inverse sensing} mechanism~\cite{lee-ieeejssc-2002, lee-isscc-2002} and \inum{ii}~internal \bxor logic~\cite{kim-ieeejssc-2012, kim-ieeejssc-2018}, \fixok{both of which} are already supported in most NAND flash chips.
Second, \passthree{we relax} \fixok{the data location \passthree{constraints} of \passthree{the} basic \mws mechanism (e.g., bitwise \bor/\bnor operations are possible only for wordlines in different NAND strings)} 
by \inum{i}~storing each operand's inverse data and \inum{ii}~leveraging De Morgan's laws.
For example, if \fixok{the} user stores the inverse of \texttt{A}, \texttt{B}, and \texttt{C} (i.e., \ol{\texttt{A}}, \ol{\texttt{B}}, and \ol{\texttt{C}}) in the same NAND string, \prop can perform bitwise \bor of three wordlines by performing bitwise \bnand of \ol{\texttt{A}}, \ol{\texttt{B}}, and \ol{\texttt{C}} 
because (\texttt{A}~\bor~\texttt{B}~\bor~\texttt{C})$=$\bnot(\ol{\texttt{A}}~\band~\ol{\texttt{B}}~\band~\ol{\texttt{C}}).

\prop requires only small changes to the control logic of a NAND flash chip, but \emph{no} changes to its cell array and sensing circuitry. 
For efficient post-fabrication tests and optimizations, most modern NAND flash chips are already capable of \inum{i}~simultaneously sensing multiple wordlines~\cite{crippa-springer-2010} and \inum{ii}~adjusting programming step and voltage \fixok{at fine granularity}~\cite{jeong-fast-2014, kim-glsvlsi-2018, luo-jsac-2016, cai-procieee-2017}. 
Hence, \fixok{integrating} \prop \fixok{ into} existing NAND flash chips requires changes only to the command latching \fixok{circuitry} and the firmware of the microcontroller in the \fixok{flash} chip (see \sect{\ref{sec:design}}).

We evaluate \prop in two ways.
First, we validate \prop using 160 real 48-layer 3D NAND flash chips. 
Our results show that \prop enables commodity NAND flash chips to perform bitwise \band/\bor/\bnand/\bnor of up to 48 operands via a single sensing operation (25~\usec).
In our validation of \fixok{computation results across} more than 10$^{11}$ flash cells, we observe \emph{zero} bit errors.
Second, we compare \prop to two different computing platforms, \fixok{a state-of-the-art} multi-core CPU \fixok{(\fixok{which we call} outside-storage processing or OSP)}~\cite{IntelCor20} and \pbit~\cite{gao-micro-2021}.
Our evaluation using three real-world workloads shows that \prop improves performance by \passthree{32$\times$/3.5$\times$} and reduces energy consumption by \passthree{95$\times$/3.3$\times$} on average compared to OSP/\pbit.

This work makes the following key contributions:
\begin{itemize}
    \item To our knowledge, this work is the first to \fixok{enable} NAND flash memory to perform \bbos on \emph{multiple} (i.e., tens) operands \js{via \fixok{a} \emph{single} sensing \fixok{operation}}.
    \item We \fixok{introduce} \prop, a new in-flash processing technique to significantly improve both performance and energy efficiency of \bbos while achieving zero \fixok{bit errors} in computation results.
    \item We demonstrate the feasibility and reliability of \prop using 160 real \fixok{state-of-the-art} 3D NAND flash chips.
    \item We evaluate the effectiveness of \prop using real-world workloads, showing large performance and energy 
    benefits over \fixok{a} state-of-the-art \fixok{multi-core processor and the state-of-the-art in-flash processing technique}.
\end{itemize}

\section{Background}\label{sec:background}
We provide a brief background on NAND flash memory that is useful to understand the rest of the paper.

\subsection{Basics of NAND Flash Memory}\label{ssec:bg_basics}
\begin{figure}[h]
 \centering
 \includegraphics[width=0.8\linewidth]{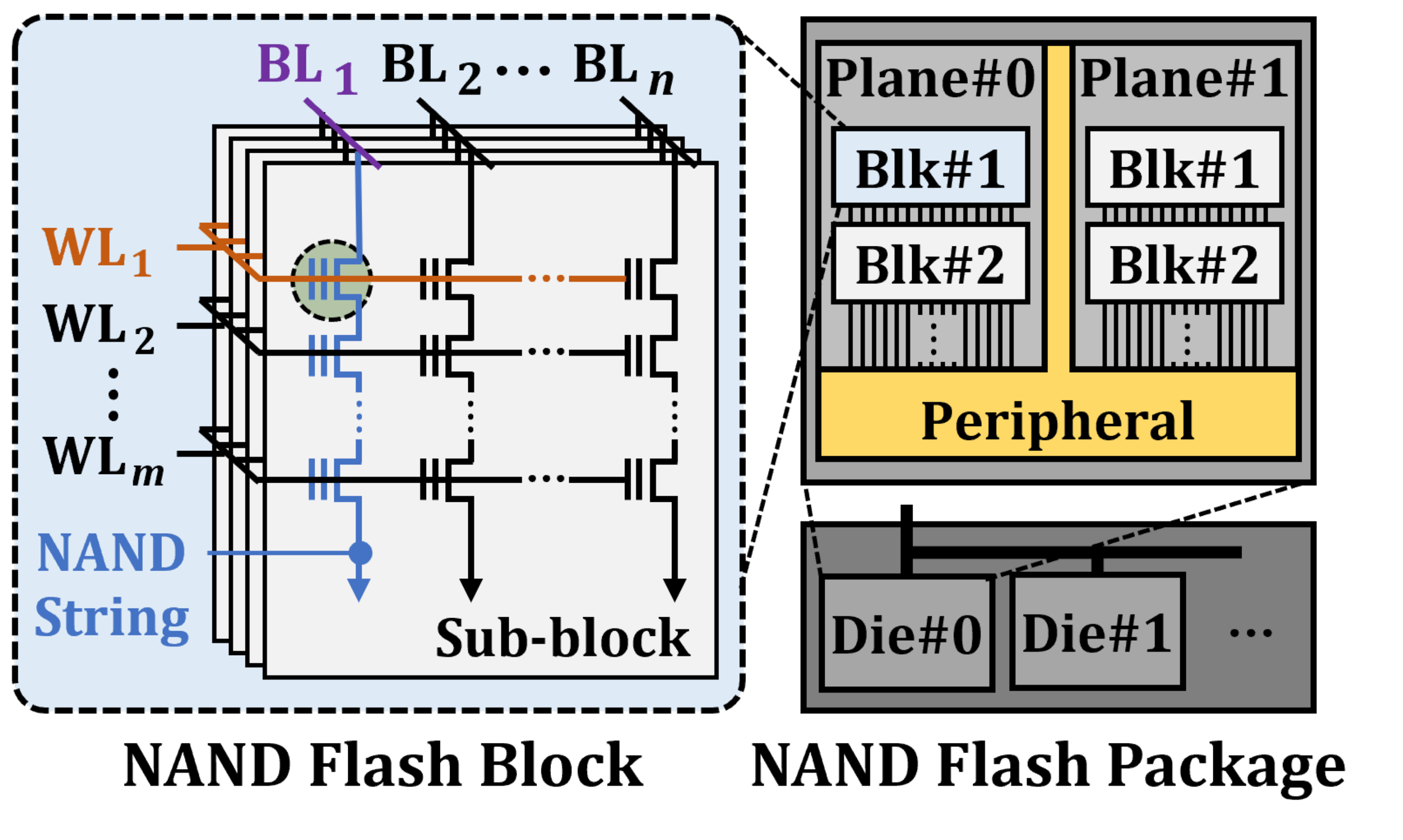}
 \caption{NAND flash organization.}\label{fig:bg_nand_organization}
\end{figure}

\head{NAND Flash Organization}
\fig{\ref{fig:bg_nand_organization}} shows the organization of 3D NAND flash memory.
A number of vertically-stacked flash cells (e.g., 24 to 176 cells) are serially connected, which is called a \emph{NAND string.}
A NAND string is connected to a bitline (BL), and 
NAND strings at different BLs compose a \emph{sub-block}.
The control gate\fixok{s} of \fixok{all} cells \fixok{that are} at the same vertical location in a sub-block \fixok{are} connected to the same wordline (WL), which makes all \fixok{such} cells operate concurrently.
A NAND flash \emph{block} consists of several (e.g., 4 or 8) sub-blocks, and thousands of blocks comprise a \emph{plane}\fixok{. The blocks in a plane share all the BLs in that plane}, which implies that a single BL is shared by thousands of NAND strings.
In the rest of the paper, unless specified otherwise, we refer to a sub-block as a block for simplicity.
A NAND flash \emph{chip} (or a \emph{die}) contains multiple (e.g., 2 or 4) \emph{planes}.
Multiple chips in a NAND flash \emph{package} can operate independently of each other but share the package's command/data buses (i.e., \emph{channel}) in a time-interleaved manner.

\head{Program and Erase Operations}
A flash cell stores data using its threshold voltage (\vth) level that highly depends on the amount of charge in the cell's charge trap.
A program operation injects electrons into a cell, which \emph{increases} the cell's \vth level.
As multiple flash cells are connected to a single WL, NAND flash memory writes data at \emph{page} \passthree{(e.g., 16 KiB)} granularity such that each cell in a WL stores one bit of the page.
To \emph{decrease} a programmed cell's \vth level, NAND flash memory performs an erase operation that ejects electrons from the cell.
The \fixok{granularity} of an erase operation is a \emph{block}, which causes the erase latency \tbers to be much longer (e.g., 3--5~ms) than the program latency \tprog (e.g., 200--700~\usec).

\begin{figure}[h]
 \centering
 \includegraphics[width=0.95\linewidth]{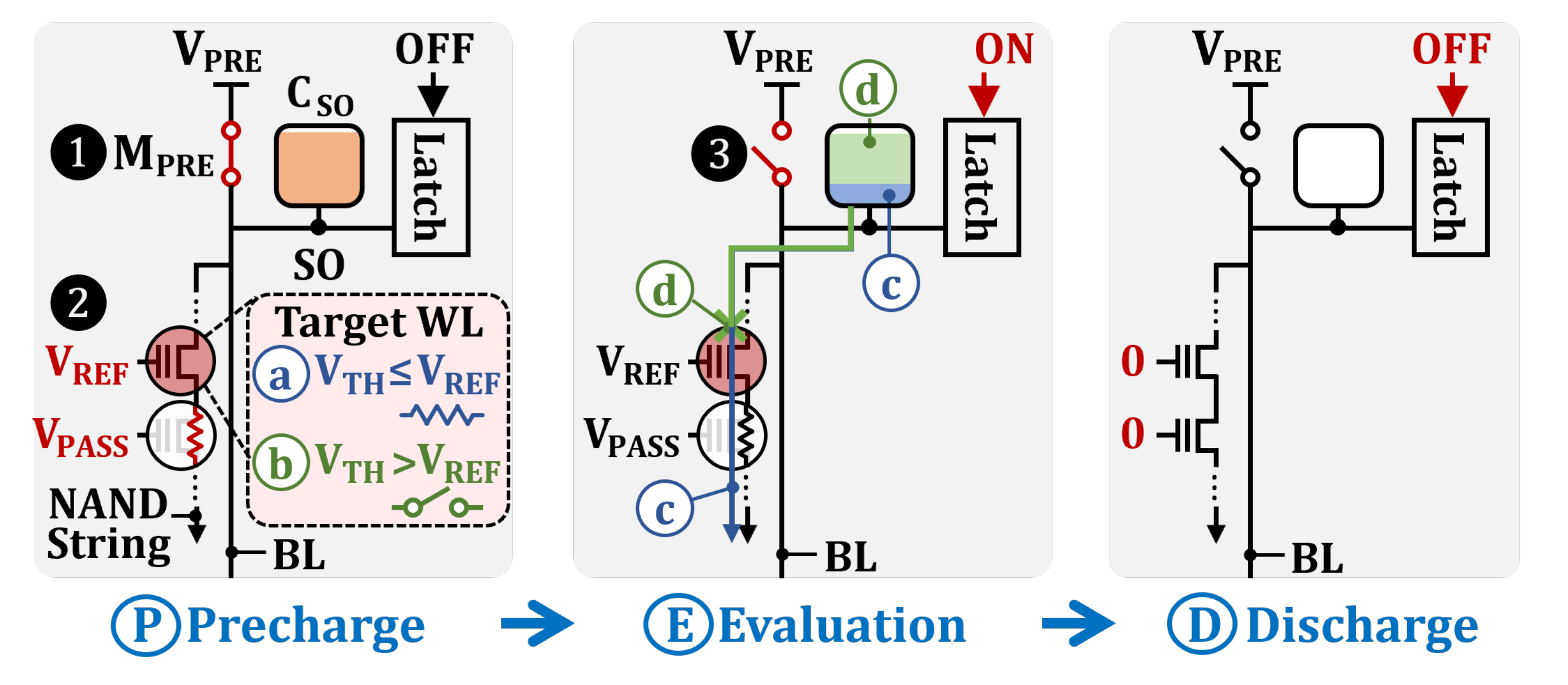}
 \caption{NAND flash read mechanism.}\label{fig:bg_read}
\end{figure}

\head{Read Operation} 
NAND flash memory determines a cell's \vth level (i.e., the cell's bit data) by sensing the conductance of the corresponding NAND string.
\fig{\ref{fig:bg_read}} shows the read mechanism of NAND flash memory\fixok{,} which consists of three steps: \inum{i}~precharge, \inum{ii}~evaluation, and \inum{iii}~discharge~\cite{micheloni-insidenand-2010, park-asplos-2021}.
In the precharge step (\wcirc{\textbf{P}} in \passthree{the left part of} \fig{\ref{fig:bg_read}}), a NAND flash chip ~charges all target BLs and their sense-out (SO) capacitors (\cso) to the precharge voltage \vpre{} \fixok{by enabling the precharge transistor M$_\text{PRE}$~\bcirc{1}}.
At the same time, the chip applies the read-reference voltage \vref to the target WL while \fixok{applying a much larger pass voltage \vpass to} the other WLs in the same block~\bcirc{2}.
Doing so makes each target cell's \vth level dictate the corresponding NAND string's conductance;
the target cell would operate as either a \fixok{resistor,} if \vthn$\leq$\vref{} (\fixok{\wcirc{\small{a}}} in \fig{\ref{fig:bg_read}, \passthree{left part}})\fixok{,} or an open switch\fixok{,} if \vthn$>$\vref (\fixok{\wcirc{\small{b}}});
\fixok{all} non-target cells in the same NAND string would always operate as \passthree{resistors} since \vpass is high enough ($>$6 V) to turn on any flash cell regardless of its \vth level~\cite{cai-dsn-2015}.
The chip then starts the evaluation of the target cells (\wcirc{\textbf{E}} in \passthree{the middle part of} \fig{\ref{fig:bg_read}}) by disconnecting the BLs from \vpre~\bcirc{3} and enabling the latching circuit~\bcirc{4}. 
If the target cell's \vth level is lower than \vref, the charge in \cso quickly flows through the NAND string \fixok{(\wcirc{\small{c}})}, which is sensed as a `\texttt{1}'.
If \vthn$>$\vref, the capacitance of \cso hardly changes \fixok{(\wcirc{\small{d}}) as the target cell blocks the BL discharge current}, which is sensed as a `\texttt{0}'.
Finally, the chip discharges the BLs (\wcirc{\textbf{D}} in \passthree{the right part of} \fig{\ref{fig:bg_read}}) to return \passfour{the} NAND string to \passfour{its} \fixok{initial \passthree{stable} state \passthree{(i.e., \passfour{the} state before precharge can take place)}} for future operations.

\begin{figure}[t]
 \centering
 \includegraphics[width=\linewidth]{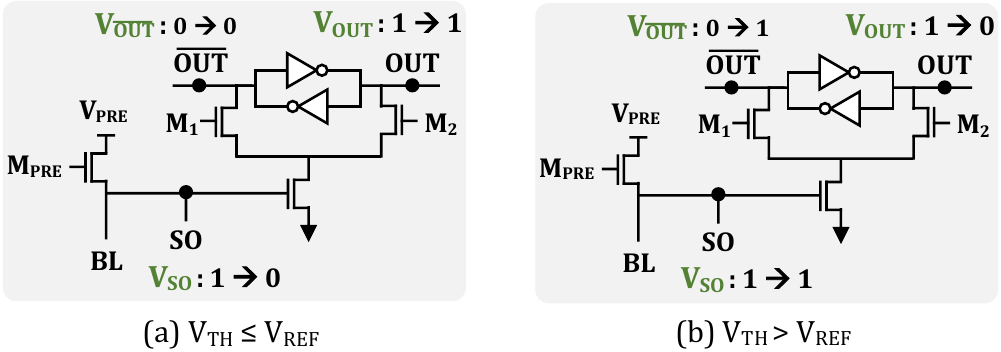}
 \caption{Latching circuit showing the voltage states \passthree{at SO, OUT and \ol{\text{OUT}}} in the precharge~\wcirc{\textbf{P}} and evaluation~\wcirc{\textbf{E}} steps during a \emph{read} operation (see \fig{\ref{fig:bg_read}}).}\label{fig:bg_latching}
\end{figure}

\fig{\ref{fig:bg_latching}} depicts how NAND flash memory senses a BL's conductance with its latching circuit.
\fixok{Figures \ref{fig:bg_latching}(a) and \ref{fig:bg_latching}(b) describe the operation of the latching circuit when the threshold voltage \vth of a flash cell is lower and higher than the read reference voltage \vref, respectively.}
\fixok{We show the transition in voltage state at each of the three nodes (SO/OUT/\ol{\text{OUT}}) when going from \passthree{the} precharge step (\wcirc{\textbf{P}} in \fig{\ref{fig:bg_read}}) to \passthree{the} evaluation step~(\wcirc{\textbf{E}} in \fig{\ref{fig:bg_read}}). During precharge, the NAND flash chip charges the BL, making V$_\text{SO}$$=$$\texttt{1}$. Before the evaluation step, the chip initializes the latching circuit by activating only transistor M$_1$, resulting in V$_{\overline{\text{OUT}}}$$=$\texttt{0} and thus V$_\text{OUT}$$=$\texttt{1}. The evaluation step disables M$_\text{PRE}$ and M$_1$ while enabling M$_2$. In Figure {\ref{fig:bg_latching}}(a), the evaluation step makes V$_\text{SO}$$=$\texttt{0} as the charge in \cso quickly flows through the NAND string (because \vthn$\leq$\vref), which leads to V$_\text{OUT}$$=$\texttt{1}~(\wcirc{\textbf{E}} in \fig{\ref{fig:bg_read}}). The bit value of the flash cell is immediately stored in the latching circuit because of the low charge retention of \cso~\cite{micheloni-insidenand-2010}. In Figure {\ref{fig:bg_latching}}(b), the evaluation step leads to V$_\text{SO}$$=$\texttt{1} and V$_\text{OUT}$$=$\texttt{0} as the flash cell operates as an open switch when \vthn$>$\vref}.

\begin{figure}[h]
 \centering
 \includegraphics[width=\linewidth]{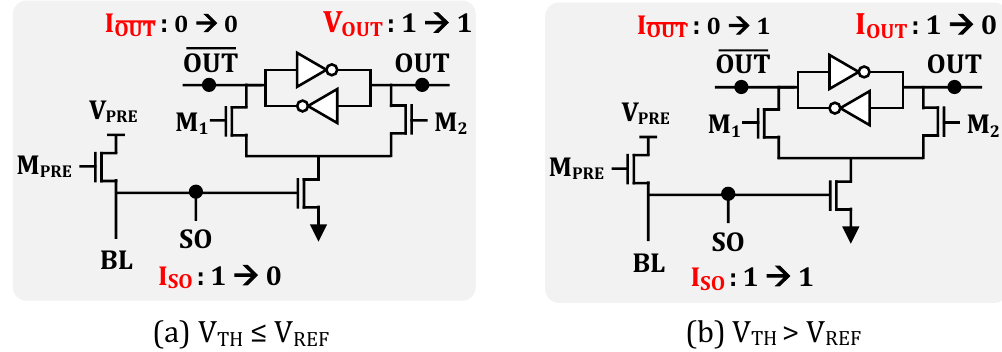}
 \caption{Latching circuit showing the voltage states \passthree{at SO, OUT and \ol{\text{OUT}}} in the precharge~\wcirc{\textbf{P}} and evaluation~\wcirc{\textbf{E}} steps during \passfour{an} \emph{inverse read}  operation (see \fig{\ref{fig:bg_read}}).}\label{fig:bg_latching_inverse_read}
\end{figure}

\head{Inverse Read} Modern NAND flash chips commonly support the \emph{inverse-read} mode~\cite{lee-ieeejssc-2002} to read the inverse of the stored data.\footnote{
Supporting inverse reads is essential to the copyback operation~\cite{lee-ieeejssc-2002} that moves a page's data to another page in the same plane without off-chip data transfer and thus can improve SSD garbage-collection performance~\cite{hong-nvmsa-2019, wu-dac-2018}.}
\fixok{
Supporting inverse reads requires no hardware changes to the latching circuit shown in \fig{\ref{fig:bg_latching_inverse_read}}.
We denote the voltage states at the three nodes (SO/OUT/\ol{\text{OUT}}) during an inverse read operation using I$_{\text{SO}}$, I$_{\text{OUT}}$ and I$_{\overline{\text{OUT}}}$. The chip performs an inverse read by simply changing the activation sequence of M$_{1}$ and M$_{2}$. Unlike a read operation, the inverse read activates M$_{2}$ \passfour{to initialize} the latching circuit before the evaluation step. This leads to I$_\text{OUT}$$=$\texttt{0} and thus
I$_{\overline{\text{OUT}}}$$=$\texttt{1}. During the evaluation step, M$_{1}$ is activated while M$_{PRE}$ and M$_{2}$ are disabled. This causes the values stored in the latching circuit after \passthree{the} evaluation step to be the inverse of the values stored in a normal read.}
\subsection{Reliability of NAND Flash Memory}\label{ssec:bg_reliability}
\begin{figure}[t]
    \centering
    \includegraphics[width=\linewidth]{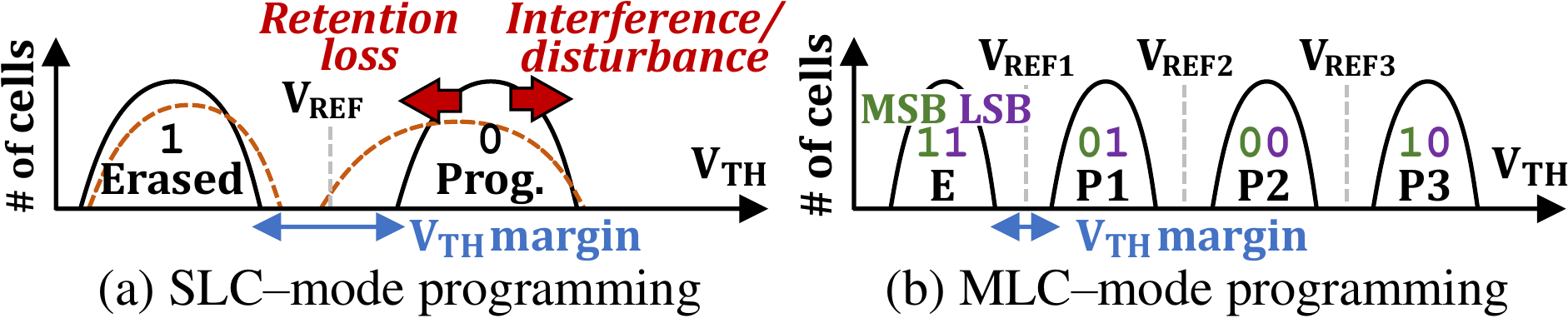}
    \caption{\vth distribution of a programmed wordline.}\label{fig:bg_vth}
\end{figure}

Modern NAND flash memory is highly error-prone due to various error sources~\cite{cai-procieee-2017} such as program interference~\cite{cai-iccd-2013, cai-hpca-2017, park-dac-2016}\fixok{,} data retention loss~\cite{cai-hpca-2015, cai-iccd-2012, cai-inteltechj-2013, luo-sigmetrics-2018}, read disturbance~\cite{cai-dsn-2015, ha-ieeetcad-2015}\fixok{, and cell-to-cell interference~\cite{cai-hpca-2017}}.
\fig{\ref{fig:bg_vth}} shows the \vth distribution of a WL, when the WL is programmed in (a) single-level cell (SLC) mode and (b) multi-level cell (MLC) mode to store one and two bits per cell, respectively.
Reading or programming a WL affects the \vth distribution of other WLs in the same block by increasing the \vth level of other cells (i.e., interference \fixok{and disturbance as shown in Figure~\ref{fig:bg_vth}(a)}). 
A flash cell also leaks its charge over time, \fixok{which decreases} its \vth level (i.e., retention loss \fixok{as shown in Figure~\ref{fig:bg_vth}(a)}).
If a cell's \vth level moves beyond \vref \fixok{ (i.e., to a \vth range corresponding to a different value)}, sensing the cell results in a different value from the original \fixok{stored} data, introducing a bit error.

Two major factors significantly increase the raw \fixok{(i.e., pre-correction)} bit-error rate (RBER) of NAND flash memory.
First, a flash cell becomes more error-prone as it experiences more program and erase (P/E) cycles~\cite{cai-date-2012}, due to the high voltage used in program and erase operations which damages the cell to more easily leak its charge.
Second, storing more bits per cell increases RBER because it reduces the margin between adjacent \vth \fixok{ ranges} \fixok{in order} to pack more \vth states within the same voltage window\fixok{,} as shown in \fig{\ref{fig:bg_vth}(b)}.

\head{Error-Correcting Codes (ECC)}
To ensure the integrity of stored data, modern SSDs commonly employ ECC.
ECC can detect and correct bit errors by storing redundant information.
To cope with the high RBER of modern NAND flash memory, it is necessary to use sophisticated ECC (e.g., low-density parity-check (LDPC) codes~\cite{zhao-fast-2013, micron-flyer-2016, cai-procieee-2017, dong-tcas-2010, zuolo-imw-2015, tanakamaru-jssc-2013, hu-fms-2012}), which increases the performance and \fixok{the} area overheads of an ECC engine.

\head{Data Randomization}
It is common practice to randomize \fixok{the values of} stored data in modern SSDs to reduce the probability of worst-case data patterns that would exacerbate program disturbance~\cite{cernea-uspatent-2006, kim-ieeejssc-2012, cha-etrij-2013}.
For example, when a NAND string has many consecutive erased cells, programming the next cell of the same NAND string significantly increases the \vth level of the consecutive cells, which \fixok{could} introduce bit errors.
Data randomization \fixok{reduces} the probability of such cases \fixok{to a} small \fixok{value} by randomly distributing \vth states across a NAND string regardless of the original data \fixok{values} to store.\footnote{In fact, \fixok{randomization} is the reason why the \vth distribution of NAND flash memory is commonly described by using \vth states with the same shape\fixok{,} as in \fig{\ref{fig:bg_vth}}.}\fixok{ The \passthree{stored} data is de-randomized during a read operation to correctly read the originally stored values \passthree{before they were randomized}.}

\section{Motivation}\label{sec:motiv}
\fixok{We describe the benefits} of in-flash processing and the main limitations of the state-of-the-art in-flash processing technique for \bbos (i.e., bitwise operations on large bit vectors)~\cite{gao-micro-2021}. 

\subsection{In-Flash Bulk Bitwise Operations}\label{ssec:motiv_parabit}
\begin{figure}[h]
    \centering
    \includegraphics[width=\linewidth]{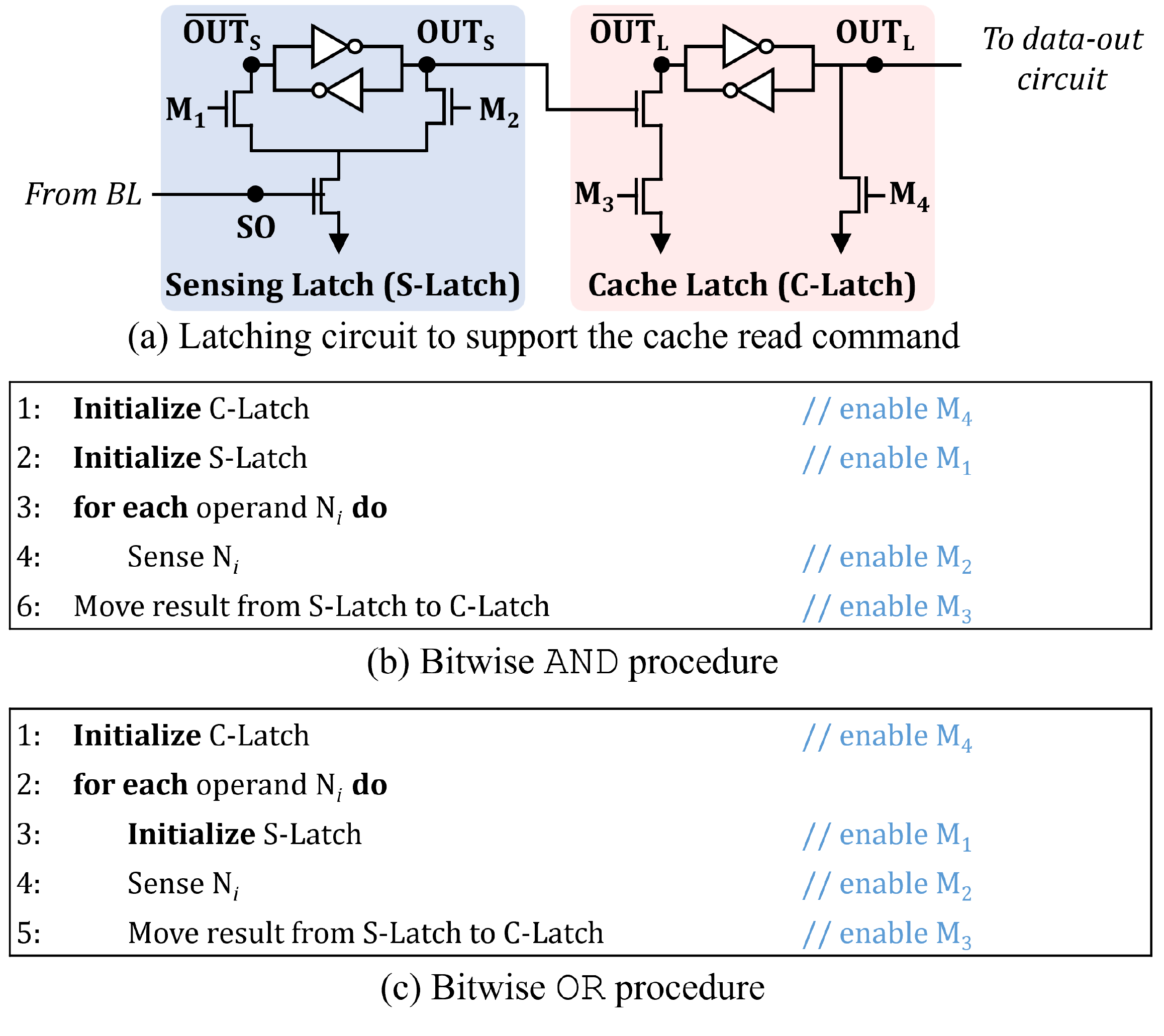}
    \vspace{-1.5em}
    \caption{Bitwise \passthree{computation techniques employed} in the state-of-the-art in-flash processing technique~\cite{gao-micro-2021}.}\label{fig:motiv_pbit}
\end{figure}

\fixok{Many prior studies~\fixme{\cite{seshadri-micro-2017, aga-hpca-2017, gao-hpca-2016, li-dac-2016, hajinazar-asplos-2021,seshadri-ieeecal-2015, seshadri-arxiv-2019}} investigate near-data processing (NDP) solutions for \bbos due to two main reasons.}
\passthree{First, \bbos are used in a \fixok{wide} variety of important applications, including databases and web search~\cite{chan-signmod-1998, oneil-ideas-2007, li-vldb-2014, li-sigmod-2013, goodwin-SIGIR-2017, seshadri-micro-2013, seshadri-micro-2017, seshadri-ieeecal-2015, hajinazar-asplos-2021, FastBitA9, wu-icde-1998, guz-ndp-2014,redis-bitmaps},
data analytics~\cite{perach-arxiv-2022, seshadri-micro-2017, jun-isca-2015, torabzadehkashi-pdp-2019, lee-ieeecal-2020},
graph processing~\cite{beamer-SC-2012, besta-micro-2021, li-dac-2016, gao-micro-2021, hajinazar-asplos-2021}, genome analysis~\cite{alser-bioinformatics-2017, loving-bioinformatics-2014, xin-bioinformatics-2015, cali-micro-2020, kim-genomics-2018, lander-nature-2001, altschul-jmb-1990, myers-jacm-1999}, cryptography~\cite{han-spie-1999, tuyls-springer-2005, manavski-spcom-2007}, set operations~\cite{besta-micro-2021, seshadri-micro-2017}, and hyper-dimensional computing~\cite{kanerva-1992, kanerva-congcomp-2009, karunaratne-nature-2020, imani-hpca-2017}}.
Second, \bbos can significantly benefit from NDP.
Due to the simple nature of bitwise operations, the performance and energy efficiency of \bbos are bottlenecked by data movement between the computation units and the memory hierarchy in \fixok{conventional systems~\cite{aga-hpca-2017, seshadri-ieeecal-2015, seshadri-micro-2017, gao-micro-2021, hajinazar-asplos-2021, seshadri-arxiv-2019}}.
NDP can effectively mitigate such data movement at low cost by supporting simple bulk bitwise operations \fixok{at \fixok{very} high \fixok{levels of} concurrency} near or inside memory devices \passthree{(e.g., in all memory banks or subarrays~\cite{seshadri-micro-2017, hajinazar-asplos-2021, kim-isca-2012})}.\\
\indent Among many existing NDP solutions, only one recent work, \pbit~\cite{gao-micro-2021}, proposes an in-flash processing technique for \bbos inside a NAND flash chip. \pbit leverages the %
latching circuits that are commonly employed in modern NAND flash chips~\cite{leong-uspatent-2008, macronix-technote-2013, micron-datasheet-2009, samsung-datasheet-2009, toshiba-datasheet-2012, micheloni-insidenand-2010} \passthree{(see \sect{\ref{ssec:bg_basics}})}.
\fixok{Existing NAND flash chips support a command called \emph{cache read}~\cite{leong-uspatent-2008, macronix-technote-2013, micron-datasheet-2009, samsung-datasheet-2009, toshiba-datasheet-2012} whose purpose is to improve the performance \passthree{of a read operation} by enabling the transfer of data from the NAND flash chip to the flash controller in parallel with the sensing of a subsequent read operation. 
To enable cache read, existing chips include a \emph{cache latch} in addition to the sensing latch \passthree{(\bcirc{4} in Figure \ref{fig:bg_read})}.}
\fixok{We describe the operation and implementation of this cache latch since it is important for and used in Parabit.}
\indent \fig{\ref{fig:motiv_pbit}(a)} illustrates the common latching circuit of a modern NAND flash chip equipped with a cache latch \passthree{(right part)} in addition to the sensing latch \passthree{(left part)} described in \fig{\ref{fig:bg_latching}}. 
A NAND flash chip initializes the cache latch in the precharge step by activating \subs{M}{4}, which pulls down node \out{L} and thus makes \nout{L}$=$\texttt{1}.
Until enabling \subs{M}{3}, the sensing latch (i.e., the value at node OUT$_\text{S}$) \emph{cannot} affect the data stored in the cache latch.
This feature enables the chip to read new data \passthree{(into the sensing latch)} \emph{while} transferring the previously-read data in the cache latch to the flash controller.

\figs{\ref{fig:motiv_pbit}(b) and \ref{fig:motiv_pbit}(c)} describe how \pbit performs in-flash bitwise \band and \bor operations, respectively, by intelligently controlling the latching circuit shown in \fig{\ref{fig:motiv_pbit}(a)}.\footnote{\pbit also introduces several mechanisms to support other bitwise operations (e.g., bitwise \bxor) and different approaches that exploit the common bit-encoding scheme for MLC NAND flash memory, but we \fixok{discuss} only \fixok{\band and \bor since} the others have key drawbacks\fixok{,} such as \passthree{costly additional inverter logic at \emph{each} BL \passfour{to} support bitwise \bxor \passfour{ operations}}.}
\linebreak
\head{Bitwise \band in \pbit}
To perform \passthree{a} bitwise \band operation, \pbit serially reads \fixok{operands sharing the same bitline} (lines 3 and 4 in \fig{\ref{fig:motiv_pbit}(b)}) while \emph{neither} enabling \subs{M}{3} \emph{nor} re-initializing the sensing latch.
Doing so allows \pbit to keep the result of \passthree{the} bitwise \band of the serially read operands in the sensing latch (node OUT$_\text{S}$).
If the read cell stores `\texttt{0}' (i.e., if the value at SO is `\texttt{1}'), enabling only \subs{M}{2} \fixok{causes} \out{S}$=$\texttt{0} \emph{regardless of} the current value at node \out{S}. 
When the cell stores `\texttt{1}' (i.e., when SO$=$\texttt{0}), sensing the cell \fixok{does} \emph{not} change the value at node \out{S}.
In other words, sensing new data N \fixok{leads} to \out{S}$=$\texttt{1} 
\emph{only if} both the new data N and the current value at node \out{S} are `\texttt{1}', which is equivalent to \out{S}$=$(N \band \out{S}).
After serially reading all the operands \fixok{sharing the same bitline} (which results in bitwise \band of all the operands at node \out{S}), \pbit enables \subs{M}{3} to move the result \passthree{from the sensing latch} to the cache latch (line 5).

\head{Bitwise \bor in \pbit} To perform \passthree{a} bitwise \bor operation, \pbit also serially reads the operands \fixok{sharing the same bitline}  (lines 2 to 5 in \fig{\ref{fig:motiv_pbit}(c)}) as in bitwise \band, but it reinitializes the sensing latch (line 3) before \passthree{sensing \emph{each} read} and activates \subs{M}{3} (line 5, \passthree{i.e., moves the result from \passfour{the} sensing latch to \passfour{the} cache latch)} after sensing \emph{each} operand. 
Doing so keeps the result of \passthree{the} bitwise \bor of the read operands \fixok{sharing the same bitline} in \fixok{the} cache latch (node \out{L}). 
If newly read data N in the sensing latch (i.e., \out{S}$=$N) is `\texttt{1}', enabling \subs{M}{3} results in \out{L}$=$\texttt{1} \emph{regardless of} the current value of node \out{L}.
When N$=$\out{S}$=$`\texttt{0}', activating \subs{M}{3} \fixok{does} \emph{not} change the value at node \out{L}.
Hence, latching new data N to the cache latch \fixok{causes} \out{L}$=$\texttt{0} \emph{only if} both the new data N and the current value of \out{L} are `\texttt{0}', which is equivalent to \out{L}$=$(N \bor \out{L}).

\begin{figure}[t]
    \centering
    \includegraphics[width=\linewidth]{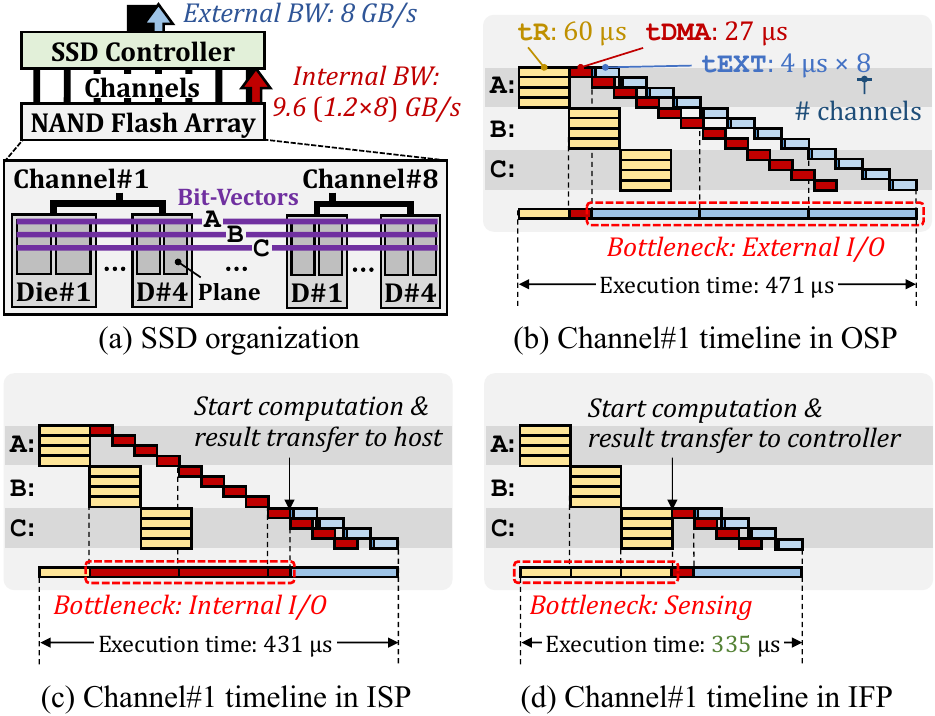}
    \caption{(a) SSD organization, and comparison of execution timelines of a channel in (b) outside-storage processing (OSP), (c) in-storage processing (ISP), and (d) in-flash processing (IFP) during \bbos.}
    \label{fig:motiv_ifp}
\end{figure}

\head{Benefits of In-Flash Processing}
\fig{\ref{fig:motiv_ifp}} shows an example where ParaBit-like in-flash processing (IFP) can provide benefits over conventional outside-storage processing (OSP) and in-storage processing (ISP) that process data using compute units \fixok{in the host CPU/GPU} and inside the SSD, respectively.
\fig{\ref{fig:motiv_ifp}(a)} depicts the target SSD considered in this example.
The SSD has eight channels, each of which is shared by four 2-plane dies (i.e., 64 planes in total) with 16-KiB pages.
We assume a page-read latency (\tr) of 60 \usec, a channel bandwidth of 1.2~GB/s between a channel and the SSD controller~\cite{kang-isscc-2019}, and an external I/O bandwidth of 8~GB/s (4-lane PCIe Gen4) between the host and \fixok{the} SSD.
\fixok{\figs{\ref{fig:motiv_ifp}(b), \ref{fig:motiv_ifp}(c) and \ref{fig:motiv_ifp}(d) show}} the execution timeline for a channel when an application uses one of OSP, ISP, and IFP, respectively, to perform \fixok{bulk} bitwise \bor operations on three 1-MiB bit vectors \texttt{A}, \texttt{B}, and~\texttt{C} \passthree{(i.e., \texttt{A} \bor \texttt{B} \bor \texttt{C})}.
We assume that each \fixok{bit-vector} is \fixok{distributed} across all the 64 planes in the SSD as shown in \fig{\ref{fig:motiv_ifp}(a)}.

Figure~\ref{fig:motiv_ifp}(b) \passthree{shows the execution timeline of \bbos for one of the eight channels in OSP}. To achieve the \fixok{highest possible} performance \fixok{using} OSP, the host must \fixok{perform concurrent multi-plane reads across the NAND flash dies for each operand. The operands themselves are read sequentially (\tr for operands A, B and C in Figure~\ref{fig:motiv_ifp}(b)). Once an operand \passthree{is} read to the sensing latch, it can be transferred to the SSD controller (\tdma) and subsequently to the host (\textr) while a read is simultaneously being performed on the next operand.} 
Given the flash channel and external I/O bandwidth, each die requires \texttt{tDMA}$=$27~\usec and \texttt{tEXT}$=$4~\usec to transfer 32-KiB data (2 planes$\times$16-KiB page) to the SSD controller and to the host, respectively.  
While \textr per die is lower than \tr and \tdma, the data movement between the host and \fixok{the} SSD \passthree{(called External I/O in Figure~\ref{fig:motiv_ifp}(b))} bottlenecks performance, as the SSD serially transfers \fixok{all the bit vectors} from the eight channels through the external I/O interface \fixok{for computation in the host CPU}.

\fixok{Figure~\ref{fig:motiv_ifp}(c) \passthree{shows the execution timeline of a channel in \passfour{the} ISP approach}}. ISP can use per-channel accelerators (e.g.,~\cite{mailthody-micro-2019, mansouri-asplos-2022, kim-fast-2021, kang-msst-2013, torabzadehkashi-pdp-2019}) to reduce external data movement by \fixok{performing computation in the SSD controller and} transferring \emph{only} the computation result to the host system.
However, the \emph{\fixok{SSD-}internal} data movement between the SSD controller and NAND flash dies \passthree{(called Internal I/O in Figure~\ref{fig:motiv_ifp}(c))} becomes the new \fixok{performance and energy} bottleneck in ISP.
This is because the internal data movement must be serialized through the channel \fixok{shared by the NAND flash dies}, while the dies \fixok{connected to} the channel can concurrently perform page reads.

\passfour{Figure~\ref{fig:motiv_ifp}(d) shows the execution timeline of a channel for the state-of-the-art IFP approach, \pbit.}
\pbit can effectively reduce both internal and external data movement \fixok{by performing the computation as the operands are read within the NAND flash chips and transferring only the computation result to the SSD controller and the host}, thereby significantly improving performance and energy efficiency. \passfour{In state-of-the-art} IFP, internal data movement between the SSD controller and the NAND flash dies is not a bottleneck, but sensing the data becomes a bottleneck, as Figure~\ref{fig:motiv_ifp}(d) shows.

\subsection{Limitations of State-of-the-Art}\label{ssec:motive_limit}

We identify two key limitations of \pbit, the state-of-the-art IFP technique for \bbos.

\head{\fixok{Unexploited Potential of IFP Capabilities}}
Despite \pbit's benefits over other processing approaches, we identify that \pbit still misses \fixok{a large potential of exploiting IFP} to significantly improve the performance and energy efficiency of \bbos.
As explained in \sect{\ref{ssec:motiv_parabit}}, \pbit \fixok{\emph{ serially}} reads every operand \fixok{from a bitline}\fixok{ (\tr in Figure~\ref{fig:motiv_ifp}(d))}.
\fixok{Each read of an operand  requires a costly (i.e., slow) sensing operation in \pbit.} \fixok{Such serial reading of every operand poses a big bottleneck when operations need to be performed across more than two operands.} 
\fixok{We} identify that NAND flash memory has \emph{inherent} capability to perform \bbos on a large number of operands \passthree{(i.e., tens)} \emph{at once} \fixok{(i.e., using only a single sensing operation)} due to \inum{i}~its unique cell-array structure and \inum{ii}~\js{flash cells' operation principles}.
First, as explained in \sect{\ref{ssec:bg_basics}}, in NAND flash memory, several tens or more than a hundred of flash cells are serially connected (as in digital \bnand gates), and thousands of NAND strings are connected to a single BL (as in digital \bnor gates).
Second, a flash cell is similar to a normal MOS transistor, a basic component for digital logic gates, in \fixok{its} structure and operation principles.
These observations lead us to develop a new IFP technique \fixok{(described in \sect{\ref{sec:idea}})} that does \passfour{\emph{not}} have the \fixok{sensing} bottleneck \passthree{that} the state-of-the-art \passfour{has: our new technique performs} \bbos \emph{on multiple operands} with \fixok{only} a \emph{single sensing operation}.

\head{Limited Applicability} \pbit's applicability is limited to \emph{highly error-tolerant} applications \passthree{because it is not designed to take into account the highly \passfour{error-prone} nature of NAND flash memory}.
As explained in \sect{\ref{ssec:bg_reliability}}, due to the error-prone nature of NAND flash memory, using ECC and data randomization is essential to guaranteeing the reliability of stored data.
However, \pbit cannot leverage any of the widely-used ECC and randomization techniques, as it performs \bbos~\emph{while} reading the operands.
Bitwise \band or \bor operation\fixok{s} on ECC-encoded or randomized data lead \fixok{can easily lead} to incorrect results \fixok{during} ECC decoding or de-randomization.
While storing fewer bits per cell can reduce RBER compared to advanced MLC techniques (e.g., triple-level cell (TLC) or quad-level cell (QLC) techniques), \pbit can be used only when the application can tolerate the NAND flash chips' RBER\passthree{ (pre-correction error rate)}\fixok{, which is \fixok{still} very large as reported in prior works~\cite{cai-procieee-2017,cai-hpca-2015,cai-hpca-2017, cai-inteltechj-2013}}. 

\begin{figure}[t]
    \centering
    \includegraphics[width=\linewidth]{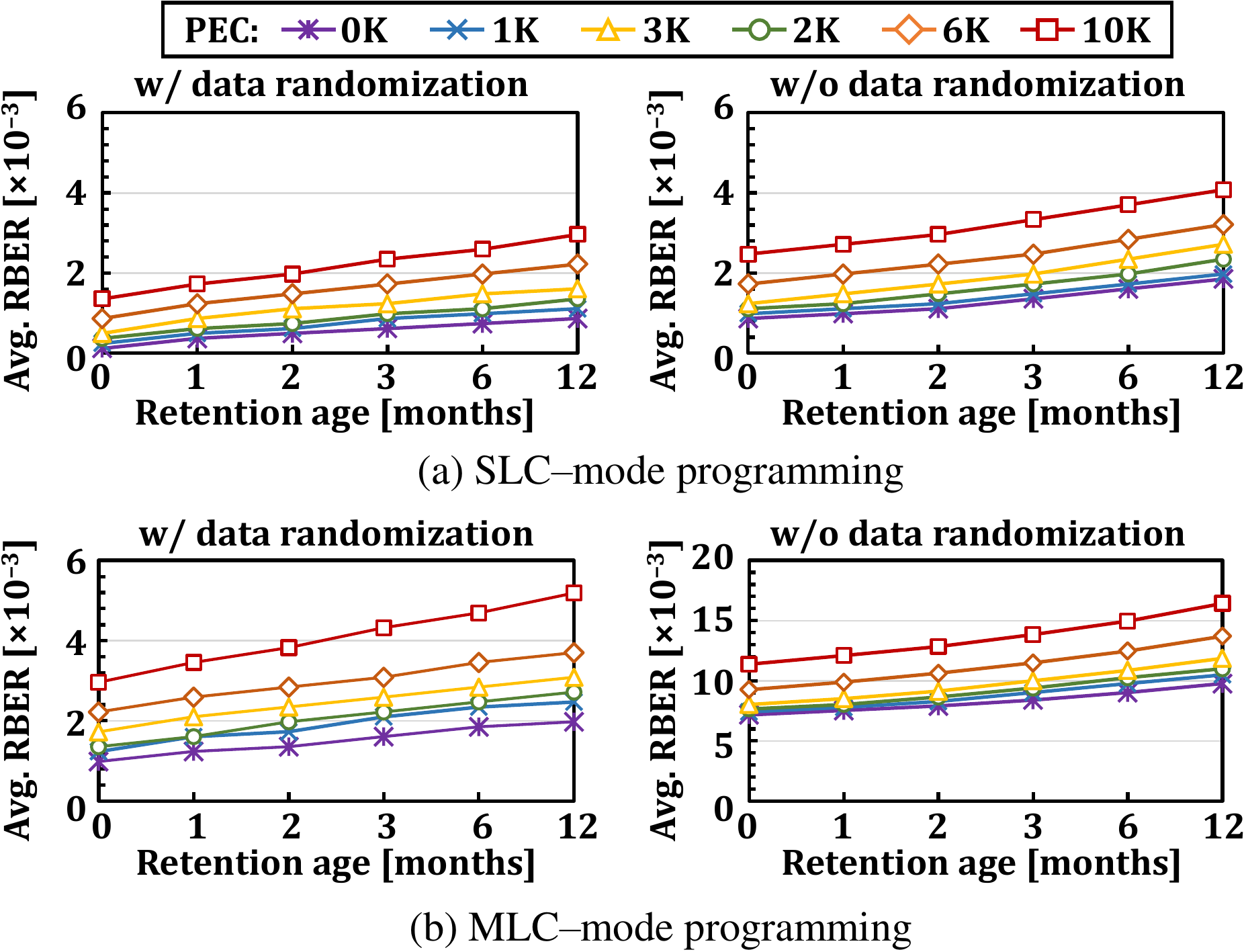}
    \caption{\passthree{Raw Bit Error Rate (RBER)} impact of \passfive{(a) SLC-mode and (b) MLC-mode} programming schemes \passthree{at different P/E cycles and retention ages, with and without data randomization}.}\label{fig:motiv_rand_rber}
\end{figure}

To better understand the impact of NAND flash reliability on the applicability of IFP, we perform real-device characterization using 160 TLC NAND flash chips (see \sect{\ref{ssec:characterization_method}} for more detail on our methodology).
\fig{\ref{fig:motiv_rand_rber}} shows the average RBER across 3,686,400 WLs randomly selected from 160 NAND flash chips \fixok{we analyze}.
We measure each WL's RBER \fixok{(without applying ECC)} when \inum{i}~programming it in (a) SLC mode and (b) MLC mode and \inum{ii}~enabling (left) and disabling (right) data randomization under different P/E-cycle counts (PEC) and retention age\fixok{s}.

We make three key observations.
First, even when using SLC-mode programming with data randomization \passthree{(left plot in Figure~\ref{fig:motiv_rand_rber}(a))}, the average RBER is significantly \passthree{(i.e., around 12 orders of magnitude)} higher than the uncorrectable bit-error rate (UBER) requirement of an SSD (e.g.,~$<$$10^{-15}$ to $10^{-16}$~\cite{gray-arxiv-2007, cox-fms-2018, cai-procieee-2017, cai-inteltechj-2013, cai-hpca-2015, cai-hpca-2017}). 
Second, disabling data randomization \passthree{(right plots in \figs{\ref{fig:motiv_rand_rber}(a) and~\ref{fig:motiv_rand_rber}(b))}} significantly increases the RBER of stored data by 1.91$\times$ and 4.92$\times$ in SLC mode and MLC mode, respectively.
Third, as expected, using MLC-mode programming \passthree{(plots in Figure~\ref{fig:motiv_rand_rber}(b))} significantly degrades the reliability of stored data, leading \fixok{to} up to 4$\times$ \fixok{the} RBER \fixok{of} SLC-mode programming.
Based on our observations, we conclude that the \fixok{state-of-the-art} \fixok{IFP} \fixok{technique} is hard to adopt for applications that cannot tolerate \fixok{a bit error rate} range of 8.6$\times10^{-4}$ to 1.6$\times10^{-2}$\passthree{ (the RBER range across the two plots in Figure~\ref{fig:motiv_rand_rber}(b))}, which is very large.

\textbf{Our goal} in this work is to develop a new in-flash processing technique that \inum{i}~\fixok{maximizes performance and energy efficiency by fully exploiting the inherent computation capability of NAND flash memory to enable many-operand \fixok{computation} with a single sensing operation} 
and \inum{ii}~provides high data reliability \fixok{(i.e., zero bit errors in computation results) so that it is applicable} to a wide range of error-intolerant applications.

\section{\prop: Key Mechanisms}\label{sec:idea}
\fixok{We} present the two key ideas of \emph{\prop (\underline{\fixok{Flash}} \underline{C}omputation with \underline{O}ne-\underline{S}hot \underline{M}ulti-\underline{O}perand \underline{S}ensing)} \fixok{that}
\js{overcome the limitations of the state-of-the-art.}

\subsection{Multi-Wordline Sensing (\mws)}\label{ssec:idea_mws}
\begin{figure}[b]
    \centering
    \includegraphics[width=\linewidth]{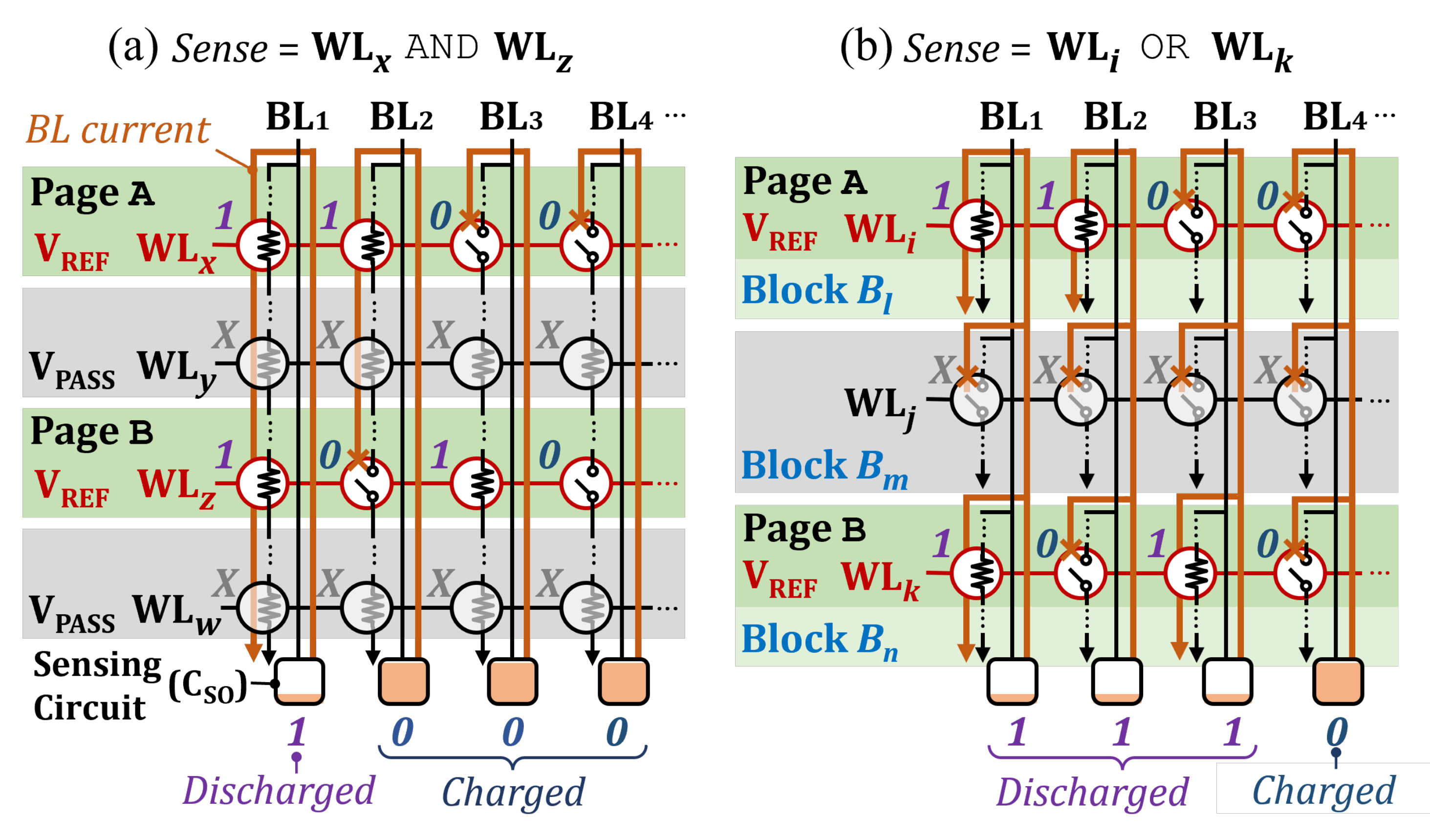}
    \caption{Overview of (a) intra-block \fixok{\mws (leading to bitwise \band)} and (b) inter-block \mws \fixok{ (leading to bitwise \bor)}.}\label{fig:idea_mws}
\end{figure}

\head{Key Idea} 
\mws is based on our key observation that simultaneously reading \fixok{\emph{multiple}} WLs in NAND flash memory results in bitwise \band or \bor of the WLs.
\fig{\ref{fig:idea_mws}} shows how \mws enables a NAND flash chip to perform bulk bitwise (a)~\band and (b)~\bor operations \fixok{on two operands} \fixok{with a} \emph{single sensing \fixok{operation}}. \fixok{While \fig{\ref{fig:idea_mws}} shows \bbos on only two operands, \prop can support \emph{multi-operand} \bbos (on tens of operands).}
\linebreak
For \textbf{bitwise \band}, the NAND flash chip simultaneously applies \vref{} to \fixok{\emph{multiple}} target WLs \fixok{that contain the source operands of the \bbo} (WL$_x$ and WL$_z$ in \fig{\ref{fig:idea_mws}(a)}) \emph{within} a block, which we call \emph{intra-block \mws}.\footnote{\passthree{Intra-block \mws , which applies \vref to two or more WLs, differs from} a regular read operation that applies \vref{} \fixok{\emph{only}} to a \fixok{\emph{single}} WL. \fixok{\pbit~\cite{gao-micro-2021}, the prior state-of-the-art IFP technique, uses regular read operations}.} 
If the chip applies \vpass to all non-target WLs (e.g., WL$_y$ and WL$_w$) as in a regular read, a BL can be \fixok{sensed as} discharged only when all the target cells in the corresponding NAND string are erased (i.e., \vthn$<$\vref{}).
In other words, the \fixok{sensing circuitry} would read a BL as `\texttt{1}' only if all the target cells store `\texttt{1}' (BL$_1$ in \fig{\ref{fig:idea_mws}(a)}) \fixok{and it would read a BL as `\texttt{0}' if any of the target cells stores `\texttt{0}'}\fixok{ (BL$_2$, BL$_3$, BL$_4$ in \fig{\ref{fig:idea_mws}(a)})}, which is equivalent to \fixok{the} bitwise \band operation. \fixok{The intra-block \mws operation can easily be generalized to more than two operands by applying \vref{} to more than two wordlines, leading to the computation of the bitwise \band of all such wordlines with a \emph{single} sensing operation.}
\linebreak
For \textbf{bitwise \bor}, \fixok{we \passthree{introduce} \emph{inter-block \mws,}}  where the chip simultaneously applies \vref{} to multiple WLs (WL$_i$ and WL$_k$ in \fig{\ref{fig:idea_mws}(b)}) of \emph{different} blocks while applying \vpass to all non-target WLs \fixok{in those blocks}.
Doing so \fixok{causes} a BL \fixok{to be} discharged if at least one of the target cells in the corresponding NAND string is erased.
In other words, the \fixok{sensing circuitry} would read a BL as `\texttt{0}' only if all the target cells \fixok{in} the BL store `\texttt{0}' (BL$_4$ in \fig{\ref{fig:idea_mws}(b)}) \fixok{and it would read a BL as `\texttt{1}' if any of the target cells stores `\texttt{1}'}\fixok{(BL$_1$, BL$_2$, BL$_3$ in \fig{\ref{fig:idea_mws}(b)})}, which is equivalent to \fixok{the} bitwise \bor operation. \fixok{The inter-block \mws operation can easily be generalized to more than two operands by applying \vref{} to more than two wordlines\passthree{, each in a different block}, leading to the computation of the bitwise \bor of \fixok{all such} wordlines across different blocks with a \emph{single} sensing operation.}

\fixok{As mentioned above,} \passthree{both types of} \mws \passthree{ are capable of} single-sensing \bbos even for more than two operands. 
This property \fixok{enables} \mws \fixok{ to be more powerful than} prior NDP proposals that leverage multi-wordline (or multi-row) activation for in-memory computation in two aspects. 
First, although prior works also propose to activate multiple wordlines to perform bulk bitwise operations inside various memory devices~\cite{aga-hpca-2017, seshadri-micro-2017, li-dac-2016, choi-iscas-2020, seshadri-ieeecal-2015}, the number of \fixok{source} operands that can be computed \fixok{on} at the same time is limited \fixok{(usually to only two)}, thereby requiring \emph{sequential} sensing of \fixok{more} operands like \pbit.\footnote{\fixok{Exceptionally, }Pinatubo~\cite{li-dac-2016} can perform \fixok{the} bitwise OR operation on a large number of rows with a single \fixok{sensing operation}. \fixok{However, Pinatubo} cannot support \fixok{the} bitwise AND operation for more than two operands.} 
Second, there exist \fixok{several} proposals that leverage multi-wordline activation to perform accumulative computation (e.g., multiply-accumulate operations) inside NAND flash memory~\cite{choi-iscas-2020, han-ieeetcas-2019, shim-ieeejetcas-2022, bayat-ieeetnnls-2017, tseng-iedm-2020, wang-ieeetvlsi-2018, lue-iedm-2019}, but they rely on analog current sensing, which requires significant changes to regular \fixok{flash chips} (for instance, \passthree{in a system with multiple memristor-based crossbar arrays, the addition of a precise analog-to-digital converter (ADC) to each array is costly (e.g., each ADC accounts for 58\% of the chip power and 31\% of the chip area even when each ADC is shared
across 128 output columns~\cite{Shafiee-isca-2016}))}.

Intra- and inter-block \mws can be combined to perform \fixok{complex} bitwise \fixok{\band and \bor} operations.\footnote{\fixok{\sect{\ref{ssec:design_device}} explains how \prop supports other common bitwise operations (e.g., \bnot, \bnand, \bnor, \bxor, and \bxnor).}}
Suppose that blocks Blk$_l$ and Blk$_n$ in \fig{\ref{fig:idea_mws}(b)} have $N$ pages (WLs), each of which stores bit vectors \texttt{A}$_i$ and \texttt{B}$_i$ (1$\leq$$i$$\leq$$N$), respectively.
If we simultaneously apply \fixok{\vref{} to} all the WLs in the two blocks, the chip would read a BL$_j$ as `\texttt{1}', only when at least one of Blk$_l$ and Blk$_n$ has \fixok{a} NAND string in which every cell stores `\texttt{1}', which is equivalent to:
\begin{equation}
     (\texttt{A}_{1,j}~\bigcdot~...~\bigcdot~\texttt{A}_{N,j})~+~(\texttt{B}_{1,j}~\bigcdot~...~\bigcdot~\texttt{B}_{N,j})
     \label{eq:combined}
\end{equation}

\head{Feasibility \& Overhead} Applying \mws to commodity NAND flash chips is highly feasible \fixok{at low cost}.
In fact, existing chips already use/support both inter- and intra-block \mws for other purposes.
For example, after erasing a block, a NAND flash chip needs to check if all the cells in the block are completely erased (\fixok{called} \emph{erase verify}) by simultaneously applying \vref{} to all the cells~\cite{micheloni-insidenand-2010}, i.e., the chip performs intra-block \mws for \fixok{\emph{all}} WLs \fixok{in the block}.
Also, manufacturers commonly design a chip to support \fixok{the} activation of multiple WLs (i.e., intra-block \mws) and multiple blocks (i.e., inter-block \mws), to perform multi-page reads/writes and multi-block erases that are \fixok{critical} for rapid testing of the chip~\cite{micheloni-insidenand-2010}.

The \mws scheme has two potential drawbacks.
First, an inter-block \mws would consume more power compared to a regular page read since it needs to activate more blocks, which requires \fixok{charging of} all the WLs in \fixok{multiple} blocks.
Note that an \emph{intra-block} \mws operation's power consumption is lower compared to a regular read because it applies \vref{} to additional target WLs, to which a regular read would apply \vpass (\fixok{which is} several times \fixok{larger} than \vref{}).
Second, the latency for a reliable \mws operation may be longer than the default read latency \tr since the target data of an \mws operation is programmed \emph{without} randomization using the \esp scheme (explained in \sect{\ref{ssec:idea_esp}}).
Without randomization, a NAND string can have low resistance (e.g., when all the cells are in the erased state) compared to \js{with} randomization (where around 50\% of the cells are \fixok{always} erased), which may increase the precharge latency and the evaluation latency \fixok{(see \fig{\ref{fig:bg_read}})} for reliable operation.
\fixok{We evaluate the potential drawbacks in \sect{\ref{sec:characterization}}}.

\subsection{Enhanced SLC-Mode Programming (ESP)}\label{ssec:idea_esp}
\begin{figure}[h]
    \centering
    \includegraphics[width=\linewidth]{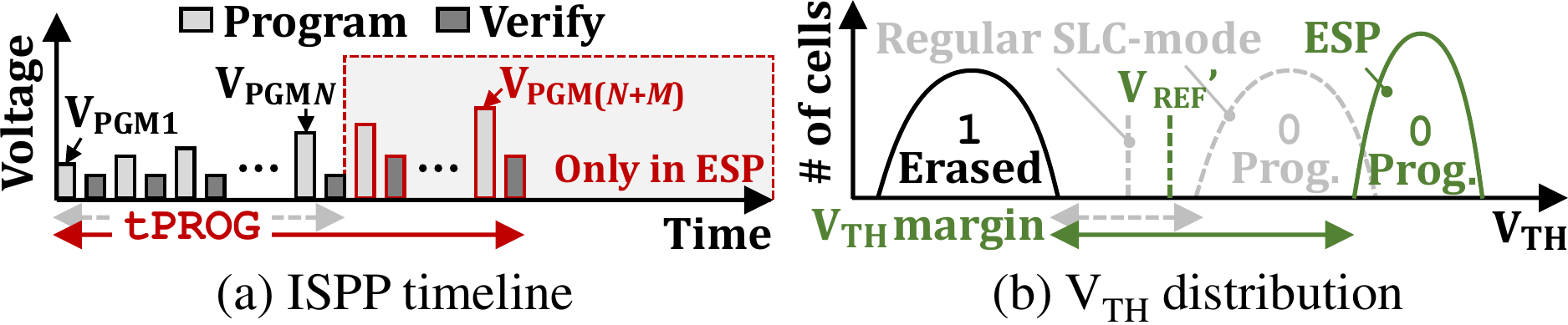}
    \caption{Overview of Enhanced SLC-mode Programming.}\label{fig:idea_esp}
\end{figure}

\head{Key Idea} \esp enhances existing SLC-mode programming by 
\js{maximizing} the margin between the two \vth states.
\fig{\ref{fig:idea_esp}} shows how ESP improves reliability compared to regular SLC-mode programming.
NAND flash memory commonly uses the \emph{incremental step pulse programming (ISPP)} scheme \cite{suh-ieeejssc-1995, jeong-fast-2014} to \fixok{precisely control the threshold voltage (\vth) of a NAND flash cell and to} narrow \fixok{the width of} \vth state \fixok{distributions}.
As \passthree{depicted} in \fig{\ref{fig:idea_esp}(a)}, the ISPP scheme gradually increases the program voltage from \vprog{1} with a certain step voltage (\vispp), until the \vth level of every cell in the WL reaches \fixok{its} target voltage \vtarget.
At the end of each ISPP step, the chip checks \js{\fixok{if} each cell\fixok{'s}} 
\vth level \fixok{has reached its} \vtarget{} (i.e.,~\textbf{Verify} in \fig{\ref{fig:idea_esp}(a)}), \fixok{and} excludes such cells \fixok{from} the next ISPP step.
\fixok{The key idea of} \esp \fixok{ is to} perform \emph{additional} ISPP steps (after performing regular SLC-mode programming), using \inum{i}~an increased \vtarget{} value \fixok{(for each cell)}, which further moves the programmed \vth state to a higher voltage level, and \inum{ii}~a decreased \vispp{} value, which narrows the width of the programmed \vth state \fixok{distribution}.
Doing so 
significantly increases the margins from the new read-reference voltage (\vref{}') to \emph{both} the erased and the programmed \vth states, as shown in \fig{\ref{fig:idea_esp}(b)}, which makes the cells less vulnerable to \fixok{many} error sources \fixok{present in NAND flash memory~\cite{cai-procieee-2017}}.

\head{Feasibility \& Overhead} 
Applying ESP to commodity NAND flash chips is highly feasible \fixok{at low cost} due to two reasons.
First, modern MLC NAND flash memory commonly supports SLC-mode programming for several reasons, e.g., storing reliability-sensitive data~\cite{kim-asplos-2020}\fixok{,} managing an SLC write buffer~\cite{lee-atc-2009} \fixok{or using unreliable cells for data storage in case they cannot be used in MLC mode~\cite{jimenez-date-2013}}.
Second, commodity NAND flash chips can tune ISPP parameters~\cite{jeong-fast-2014, shim-micro-2019, kim-glsvlsi-2018, feng-iccd-2017, Wang-ieeetcom-2016, dong-ieeetcas-2010} using \fixok{the} \texttt{SET~FEATURE} command~\cite{onfi-2020, park-asplos-2021}, which is essential to post-fabrication optimization of NAND flash chips. \fixok{The \texttt{SET~FEATURE} command is also used to dynamically adapt to changes in P/E cycles\passthree{~\cite{cai-insidessd-2018, cai-procieee-2017}} and reliability characteristics of NAND flash cells\passthree{~\cite{cai-insidessd-2018, cai-procieee-2017}}.}

\esp has two drawbacks.
First, it increases the program latency by performing additional ISPP steps.
Second, because it uses SLC-mode \passfour{programming,} it requires \fixok{double the} WLs to store the same amount of data compared to MLC-mode programming. We provide a detailed analysis of the performance and capacity overheads of \esp in \sect{\ref{ssec:overhead}}.

\begin{figure*}[t]
    \centering
    \includegraphics[width=\textwidth]{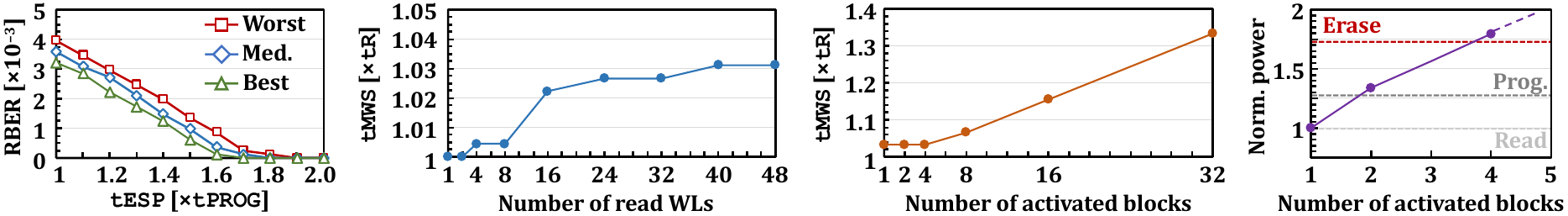}
    \begin{minipage}[t]{.22\linewidth}\vspace{-.5em}\centering\caption{RBER vs. \tesp.}\vspace{-.5em}\label{fig:characterization_esp}\end{minipage}\hfill
    \begin{minipage}[t]{.29\linewidth}\vspace{-.5em}\centering\caption{Intra-block \mws latency.}\vspace{-.5em}\label{fig:characterization_intra_mws}\end{minipage}\hfill
    \begin{minipage}[t]{.29\linewidth}\vspace{-.5em}\centering\caption{Inter-block \mws latency.}\vspace{-.5em}\label{fig:characterization_inter_mws}\end{minipage}\hfill
    \begin{minipage}[t]{.19\linewidth}\vspace{-.5em}\centering\caption{MWS power.}\vspace{-.5em}\label{fig:characterization_power}\end{minipage}
 \end{figure*}
 
\section{Real Device Characterization}\label{sec:characterization}
This section presents our characterization of \fixok{160} real 3D NAND flash chips to validate the feasibility, performance, and reliability of the two key mechanisms \fixok{of} \prop.

\subsection{Characterization Methodology}\label{ssec:characterization_method}
\head{Infrastructure}
We use an FPGA-based testing platform that contains a custom \fixok{NAND} flash controller and a temperature controller.
The flash controller supports all the commands implemented in our NAND flash chips, \fixok{including} not only basic read/program/erase commands but also various test-mode commands necessary \fixok{to dynamically} change operating parameters (e.g., the ISPP step voltage and other timing parameters) and \fixok{simultaneously activate} multiple WLs.
The temperature controller maintains a NAND flash chip within \textpm\degreec{1} of the target temperature.
This feature allows us to \inum{i} test all the chips under the same operating temperature (\degreec{85}) and \inum{ii} accelerate retention loss based on Arrhenius's Law~\cite{arrhenius-zpc-1889}\fixok{, which is essential to real device characterization under long retention ages (e.g., 1 year) \passthree{while maintaining reasonable testing} time}.
We characterize 160 48-layer 3D TLC NAND flash chips\fixok{, where} a NAND string consists of 48 flash cells \fixok{and} the page size is 16 KiB.
Under regular SLC-mode programming, the chips \passthree{have a} read latency \texttt{tR}$=$22.5 \usec and a program latency \texttt{tPROG}$=$200 \hypertarget{SC6c1}{\usec}.

\head{Methodology}
To minimize the potential \fixok{inaccuracies} in our characterization results,
we carefully design our experiments following the JEDEC standards~\cite{jedec-stress-2010, jedec-electrically-2009} that specify the test methodology recommended for evaluating the reliability of commercial-grade NAND flash products.
To ensure \fixok{high-}confidence reliability tests, \fixok{JEDEC} recommends testing more than 39 flash chips from three different wafers. 
Our 160 flash chips are fabricated from five \fixok{different} wafers, and we select 120 blocks (not sub-blocks) from each of the 160 chips at random locations.
We test every page in each selected block (\fixok{a total of} 3,686,400 WLs) to obtain statistically significant results. 

To evaluate \prop' reliability under the worst-case operating conditions, we measure each WL's RBER under a 1-year retention age at \degreec{30}~\cite{cox-fms-2018} and 10K P/E cycles.\footnote{In our RBER measurements, we exclude \emph{faulty} cells that introduce permanent (non-transient) errors due to defects in fabrication since \inum{i}~faulty cells can be profiled and excluded for the purpose of \prop, and \inum{ii}~manufacturers can significantly reduce the faulty-cell \fixok{fraction} by specially designing chips for \prop at the cost of yield and/or technology node \fixok{size}.}
We increase a block's P/E-cycle count by repeating the cycle of programming every page in the block (in TLC mode) and erasing the \fixok{entire} block.
Unless specified otherwise, we program each page using the \emph{checkered} data pattern, the worst-case data pattern for NAND flash reliability where any two adjacent cells (both horizontally and vertically) \fixok{are} programmed either \fixok{to} the highest \vth state (e.g., \passthree{the} P7 state in TLC mode) or \fixok{to} the lowest \vth state (i.e., \passthree{the} Erased state).

\subsection{Characterization Results}\label{ssec:characterization_results}
\head{ESP Latency \& Reliability}
We first study the trade-off between the reliability and program latency of the \esp scheme.
\fig{\ref{fig:characterization_esp}} shows the average RBER per 1-KiB data when we program the WLs while increasing the \esp latency \tesp for more precise ISPP control.
We plot the RBER of the worst, median, and best block out of \fixok{all} the tested blocks and normalize the increased \tesp values to the default program latency \tprog for regular SLC-mode programming.

We make two key observations from \fig{\ref{fig:characterization_esp}}.
First, it is possible to \js{avoid raw bit errors}  
without data randomization by enhancing SLC-mode programming, at the cost of \fixok{an} increase in program latency.\footnote{We also tried to achieve the same level of RBER by enhancing MLC-mode programming, but \passthree{the} RBER \passthree{of MLC-mode programming} does not decrease below 10$^{-4}$ even when \fixok{we} increase the program latency to 5~ms.}
When we increase \tesp by more than 90\% compared to \tprog, we observe \emph{zero bit errors} in our tested pages that contain more than 4.83$\times$10$^{11}$~bits \hypertarget{SC6c2}{in} total, which means that the statistical RBER of ESP is lower than 2.07×10$^{-12}$.\footnote{The result  is not enough to \emph{guarantee} the \passthree{JEDEC-specified} UBER requirements (see ~\sect{\ref{ssec:characterization_method}}), but we carefully select the evaluated blocks to avoid \fixok{any} potential \fixok{inaccuracy in} our results. Given the limitations in academia, it is challenging to experimentally guarantee the UBER requirements, which requires around 1,000$\times$ samples and \fixok{1,000$\times$ longer testing} time \passthree{than our experiments} \fixok{using} our \passfour{experimental} \fixok{testing} infrastructure.}
Second, the \esp scheme's reliability improvement significantly increases with \tesp.
For the median block, increasing \tesp by 60\% achieves an order of magnitude RBER reduction.
We conclude that the \esp scheme is essential for achieving effectively zero \passthree{bit} errors in \passthree{the computation results of} in-flash processing and \passthree{thus \esp} can increase applicability of in-flash processing to a \passthree{much} wider range of applications.  

\head{MWS Latency \& Reliability}
We measure \tmws, the minimum latency for \passthree{an} \mws operation to achieve zero bit errors in all the tested blocks.
First, we perform intra-block \mws operations while changing the number of read WLs from 1 to 48 (the number of WLs in a \js{NAND string}).
Second, we perform inter-block \mws operations on all WLs in the target blocks while changing the number of activated \passthree{(target)} blocks from 1 to 32.
In both experiments, we first increase a target block's P/E cycle count using a checkered data pattern. Next, we check the correctness of intra-block \mws \passfive{ under high disturbance and noise, induced} by programming the block using a different data pattern. This new data pattern maximizes the resistance of NAND strings in the block by programming the block to meet two conditions; \inum{i}~the number of cells that store bit value `\texttt{1}' (i.e., `\texttt{1}' cells) must be less than two; \inum{ii}~if a NAND string has a `\texttt{1}' cell, the cell must be in one of the \mws operations' target wordlines. As explained in~\sect{\ref{ssec:idea_mws}}, bypassing data randomization can affect the read latency (both \tr and \tmws) for reliable operation by decreasing a NAND string's resistance compared to the typical read of randomized data.
We validate the correctness of \mws by comparing the result of \mws operations that we obtain from the real chips to the correct results of the bitwise operations on the stored data.

\figs{\ref{fig:characterization_intra_mws} and \ref{fig:characterization_inter_mws}} show the \tmws value \fixok{(as a multiple of \tr)} for intra- and inter-block \mws operations, respectively.
We make three key observations from the results.
First, bypassing data randomization does \emph{not} increase a regular read operation's latency.
As shown \fixok{in} \fig{\ref{fig:characterization_intra_mws}}, using the default read latency introduces no error when we read only a single WL in a block.
Second, intra-block \mws does not significantly increase the read latency.
Even when \fixok{we} simultaneously read all the \fixok{48} WLs in a block, \tmws is only 3.3\% higher than \tr.
When we perform intra-block \mws on eight (or \fixok{fewer}) WLs, \tmws \fixok{ is less than 1\% higher than} \tr.
Third, although inter-block \mws \fixok{ (shown in \fig{\ref{fig:characterization_inter_mws}})} affects \tmws more significantly compared to intra-block \mws, it can still provide significant benefits over individual reads of the same WLs. 
As shown in \fig{\ref{fig:characterization_inter_mws}}, when simultaneously reading WLs in 32 different blocks, \tmws \fixok{ is} 36.3\% higher than \tr.
This is because activating multiple blocks significantly increases the number of WLs to precharge at the same time.
The increased WL-precharge time is \fixok{mostly} hidden by the BL-precharge time until we activate eight blocks, but \fixok{it becomes larger than the BL-precharge time} as the number of activated blocks further increases, \fixok{which causes the latency of a reliable \mws operation to be longer than the latency of a regular page read}. 
However, the increased latency \fixok{of \mws on 32 WLs } (1.363$\times$\tr) is much lower than the latency to individually \fixok{(serially)} read 32 WLs
(32$\times$\tr).

Based on our observations, we draw \fixok{three major} conclusions.
\fixok{First, we demonstrate that real commodity NAND flash chips can reliably support both intra-block and inter-block \mws operations, so computer architects can build a system that leverages \prop, as long as they have access to the command interfaces used for our characterization.
Second, both \passthree{types of} \mws significantly accelerate in-flash \bbos in commodity NAND flash chips at \fixok{low} cost.
Third, it is possible to support both \passthree{types of} \mws with a \fixok{small} latency increase over the default read latency.}
If we limit the maximum number of simultaneously-activated blocks for inter-block \mws to 4, we can support any \mws operation with a fixed latency (\tmws) only 3.3\% higher than \tr. 

\head{\fixok{Maximum} Power Consumption of Inter-Block MWS}
As explained in \sect{\ref{ssec:idea_mws}}, inter-block \mws consumes more power compared to a regular page read due to the \fixok{higher} number of activated WLs at the same time.
Understanding the impact \fixok{of \mws} on power \fixok{consumption} is important because a NAND flash-based SSD has a limited power budget (e.g., 75W for PCIe Gen4 SSDs\fixme{~\cite{pcie-2017}}).
\fig{\ref{fig:characterization_power}} shows the average power consumption of a NAND flash chip when we perform inter-block \mws \fixok{ as a function of} the number of \fixok{simultaneously-activated} blocks.
To measure the worst-case power consumption, we read only one WL per each block (i.e., we apply \vref{} to only one WL per block while applying \vpass ($>$\vref) to all non-target WLs). 
We normalized all values in \fig{\ref{fig:characterization_power}} to the average power consumption \passthree{of} a regular page-read operation.

We make three observations.
First, the power consumption of a NAND flash chip considerably increases with the number of activated blocks for inter-block \mws.
\passthree{Increasing} the number of activated blocks \passfour{from one to two} increases the average power consumption by about 34\%.
Second, despite the non-trivial increase in power consumption, it is possible to support inter-block \mws within the SSD's power budget.
Until we activate four blocks, the power consumption of inter-block \mws remains lower than that of an erase operation.
Third, inter-block \mws is more \emph{energy efficient} compared to serial reads of the same WLs.
For example, performing an inter-block \mws operation on four different blocks would cause about 80\% power increase compared a regular read, but due to its negligible latency increase (3.3\%), it significantly reduces the energy consumption by 53\% compared to individual reads of the four WLs.
We conclude that, with a proper limit \fixok{on the number of} inter-block \mws,
\prop would not require an increase in the power budget of commodity SSDs.
\section{Design of \prop}\label{sec:design}
\fixok{We present} our design of \prop to support efficient in-flash \bbos.

\subsection{\fixok{Enhanced Computation Capability}}\label{ssec:design_device}
We enhance the \fixok{basic} capability of \prop \fixok{ (beyond the bitwise \band and bitwise \bor operations introduced in ~\sect{\ref{ssec:motiv_parabit}})} in two ways.

\head{Supporting Other Bitwise Operations} 
We design \prop to \fixok{also} support bitwise \bnot/\bnand/\bnor/\bxor/\bxnor operations by leveraging two existing features that are  widely supported in \fixok{real} NAND flash memory \fixok{chips}.
First, as explained in~\sect{\ref{ssec:bg_basics}}, modern NAND flash memory commonly supports inverse reads~\cite{lee-ieeejssc-2002}, which enables \prop to perform not only bitwise \bnot operations but also bitwise \bnand and \bnor operations. 
\fixok{\prop performs \bnot of a WL by simply reading the WL in inverse-read mode.}
If we perform intra-block (inter-block) \mws while controlling the sensing latch circuit in inverse-read mode, \fixok{the sensed data would be} the inverse value of \fixok{the} bitwise \band (\bor) of all the WLs simultaneously read, i.e., bitwise \bnand (\bnor) by definition.

Second, many modern NAND flash chips (including the \fixok{160} chips used in our real-device characterization, \fixok{~\sect{\ref{sec:characterization}}}) support a bitwise \bxor operation between the data in different latches \fixok{(i.e., two/three additional latches available in a NAND flash chip for MLC/TLC program operation)}~\cite{kim-ieeejssc-2012, kim-ieeejssc-2018}. 
This feature is essential for supporting on-chip randomization~\cite{kim-ieeejssc-2012} and improving testability (e.g., it significantly reduces a NAND flash chip's test time by enabling \fixok{comparison} of programmed data \fixok{to a golden value} without reading the data out of the chip~\cite{lu-jet-2018,cao-elecletters-2022}).
By using this feature along with inverse reads, \prop can also support bitwise \bxnor operations since
\begin{equation}
    \bxnorxy{A}{B} \equiv \bxorxy{\ol{\texttt{A}}}{B} \equiv \bxorxy{A}{\ol{\texttt{B}}}
\end{equation}  
To be specific, \prop uses the existing \bxor logic while performing an inverse read for either of the two operands.

\head{\passthree{Improving the Performance of \passfour{the} Bitwise \bor Operation}}
\passthree{The performance of bitwise \bor using inter-block \mws is limited compared to that achieved by bitwise \band using intra-block \mws.
As demonstrated in \sect{\ref{ssec:characterization_results}}, commodity NAND flash chips can perform bitwise \band of all the WLs in a block via a \emph{single} intra-block \mws operation. 
However, the maximum number of operands in inter-block \mws is limited due to high power consumption.

We can remove this restriction on the maximum number of activated blocks by performing bitwise \bor using 1) intra-block \mws along with 2) \emph{inverse reads} and \passfive{3) taking advantage of \passsix{De Morgan's} laws.}}
If we store operands in a block with their \emph{inverse} data (instead of the original data), we can perform bitwise \bor of the operands with a single \emph{intra-block} \mws operation %
by leveraging the inverse read mode and De Morgan's laws:
\begin{equation}
     (\texttt{A}_{1}~+~...~+~\texttt{A}_{N})\equiv\overline{(\overline{\texttt{A}_{1}}~\bigcdot~...~\bigcdot~\overline{\texttt{A}_{N}})}
\end{equation}  
Note that \inum{i}~\prop can return the original data of such operands via inverse reads ($\because$\texttt{A}$\equiv$\bnot\ol{\texttt{A}}), and \inum{ii}~inter-block \mws is still useful for \fixok{combined bitwise \band/\bor operations as explained in \sect{\ref{ssec:idea_mws}} (Equation~\ref{eq:combined})}.

\head{Increasing Maximum Number of Operands for IFP}
\prop alone \emph{cannot} completely avoid off-chip data transfer for bitwise \band/\bor/\bnand/\bnor operations if the number of operands exceeds the number of WLs in a block.
\fixok{This is because intra-block \mws involves a single sensing operation to read all WLs within a block, thereby limiting the maximum number of operands to the number of WLs in a block.} Inter-block MWS has a stronger constraint on the maximum number of operands due to the limited power budget as explained in \sect{\ref{ssec:characterization_results}}. 

Fortunately, \prop can \emph{accumulate} the results of multiple intra-block MWS operations by leveraging \pbit, which has \passfour{fewer} constraints on the maximum number of operands. 
For example, suppose that \inum{i}~a block has $N$ WLs, and \inum{ii} \prop needs to perform bitwise \band of all WLs of $M$ different blocks (i.e., the number of total operands is $M\times{N}$). We can accumulate the results in two steps. First, \prop performs bitwise \band on each block for $N$ operands at a time.
\passfive{Second, \prop performs bitwise \band on the results from the $M$ blocks.}

\subsection{\prop Command Set}\label{ssec:design_cmd}
Although we demonstrate that our NAND flash chips already support all necessary features to perform the \esp and \mws operations in their test-mode command set, efficient design of \prop commands is important due to two reasons.
First, NAND flash vendors consider their test-mode command set design to be proprietary and do not reveal any details in publicly-accessible documentation.
Second, efficient command set design can significantly reduce the necessary changes to the NAND flash chip's control logic and communication overheads with a flash controller.

\begin{figure}[h]
    \centering
    \includegraphics[width=\linewidth]{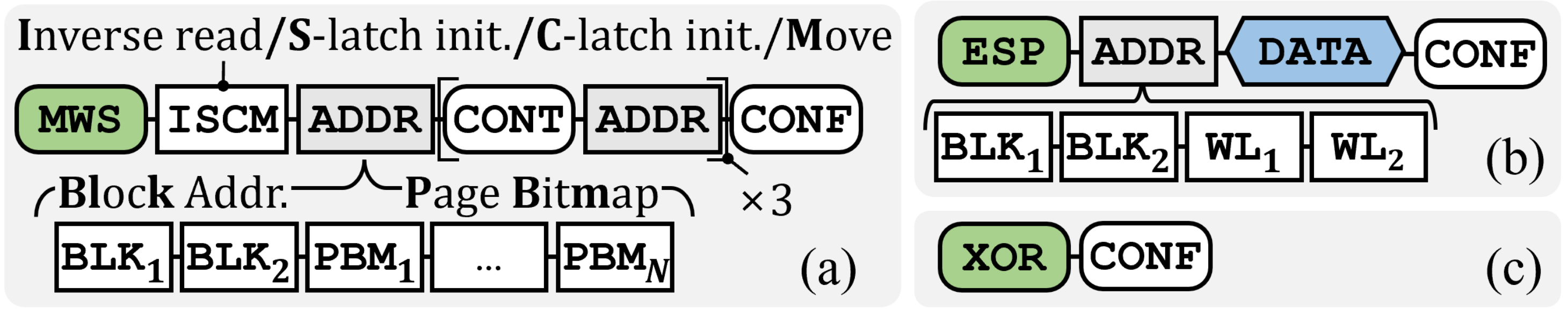}
    \caption{\passthree{Three new NAND flash commands for \prop: (a)~\texttt{MWS}, (b)~\texttt{ESP}, and ~\texttt{XOR}}.}\label{fig:design_cmd}
\end{figure}

\fig{\ref{fig:design_cmd}} shows three new NAND flash commands that we design for \prop: (a)~\texttt{MWS}, (b)~\texttt{ESP}, and (c)~\texttt{XOR}.
We design the \texttt{MWS} command to be used for all three necessary features in \prop except for bitwise \bxor: \inum{i}~intra- and inter-block \mws, \inum{ii}~inverse read, and \inum{iii}~accumulation of the results of all reads as  described in \sect{\ref{ssec:design_device}}.
To this end, we extend the regular read command that contains the operation code and target page address in three aspects.
First, we add the \texttt{ISCM} command slot before the address slot to allow a flash controller to turn on/off four features by setting the dedicated flags: \inum{i}~inverse-read mode, \inum{ii}~sensing-latch (S-latch) initialization, \inum{iii}~cache-latch (C-latch) initialization, and \inum{iv}~data transfer from S-latch to C-latch.
Second, we enable the flash controller to efficiently specify the WLs to activate for \mws operations by sending the page bitmap (PBM) instead of the page index in the address slot.~Third, we design an \mwsc command to have up to four address slots for inter-block \mws by sending the additional block address and PBM after \fixok{a} \texttt{CONT} (continue) slot.\fixok{\footnote{\texttt{CONT} is a command slot to indicate that an address cycle will follow next. \texttt{CONF} is a command slot to indicate the end of the command sequence.}} 
The \espc command has the same command interface as the regular program command, and the \bxor command performs bitwise \bxor between two (sensing and cache) latches and store\fixok{s} the result in the C-latch.

\begin{figure}[t]
    \centering
    \includegraphics[width=\linewidth]{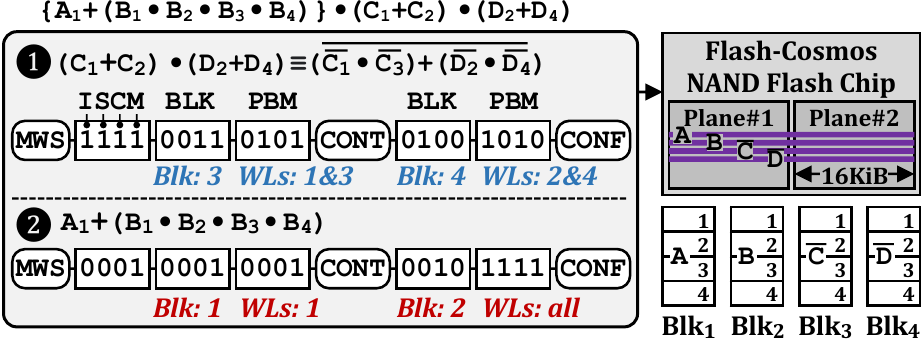}
    \caption{Operational example of \prop.\passfive{While performing \passsix{the second \mws command}, the results of the two \mwsc commands, \bcirc{1} and \bcirc{2}, are accumulated in both \passfive{the} S-latch and \passfive{the} C-latch.}}\label{fig:design_ex}
\end{figure}

\fig{\ref{fig:design_ex}} shows how a flash controller can use \prop to perform \bbos \fixok{ using} an example \fixok{that makes} two assumptions\fixok{:}
\inum{i}~a \prop-enabled chip stores four sets of four bit vectors, A$_i$, B$_i$, C$_i$, and D$_i$, in four blocks Blk$_{i}$ (1$\leq$$i$$\leq$4), each of which has four pages;
\inum{ii}~bit vectors C$_i$ and D$_i$ are programmed using their inverse data with the knowledge that they would be used for bitwise \bor \passthree{ (\sect{\ref{ssec:design_device}})}.
Suppose that the user \passthree{would like} to perform the following bitwise operations:
\begin{equation}
     \{\texttt{A}_1+(\texttt{B}_{1}\bigcdot\texttt{B}_{2}\bigcdot\texttt{B}_{3}\bigcdot\texttt{B}_{4})\}\bigcdot(\texttt{C}_{1}+\texttt{C}_{3})\bigcdot(\texttt{D}_2+\texttt{D}_4)
\end{equation}
As shown in \fig{\ref{fig:design_ex}}, the user can perform bitwise operations using two \fixok{intra-block} \mwsc commands, \bcirc{1} one for $(\texttt{C}_{1}+\texttt{C}_{3})\bigcdot(\texttt{D}_2+\texttt{D}_4)$ while enabling the inverse-read mode and initialization of both latches and \bcirc{2} the other for $\texttt{A}_1+(\texttt{B}_{1}\bigcdot\texttt{B}_{2}\bigcdot\texttt{B}_{3}\bigcdot\texttt{B}_{4})$ while disabling the inverse-read mode and initialization of both latches. \passthree{By disabling the initialization of both latches while performing \passsix{the second \mws command}, \passfive{the results of the two \mwsc commands, \bcirc{1} and \bcirc{2}, }are accumulated in both \passfive{the} S-latch and \passfive{the} C-latch (\sect{\ref{ssec:design_device}}).}
Note that the order of the two \mwsc commands is important, as an inverse read requires S-latch initialization\fixok{, which prevents the accumulation of the results}.

\subsection{\fixok{System Support}}\label{ssec:design_ssd}
\fixok{We briefly} discuss the end-to-end system \fixok{support} that we envision to efficiently \fixok{enable} \prop.

\head{Requirements}
There are two key requirements for a system to take full advantage of \prop-enabled NAND flash chips. 
First, the target data of bitwise operations needs to be properly stored using \esp. 
To maximize the performance benefits of \prop, it is important to store as many operands of the target bitwise operation as possible in the same block, which minimizes the number of \mws operations required. 
For example, bitwise \bor on 48 pages (i.e., operands) would require 12 inter-block \mws operations if each operand is stored in different blocks, assuming that the maximum number of pages for inter-block \mws is limited to 4 in order to avoid \fixok{maximum} power \fixok{consumption related} issues.
However, when the operands are stored in the same block with their inverse data, it is possible to perform the same bitwise \bor operation \fixok{with} a single intra-block \mws operation \fixok{using} inverse-mode read \passthree{(\sect{\ref{ssec:design_device}})}. 
Second, the host system needs to interact with the underlying SSD in order to \fixok{efficiently store the data to maximize the benefits of \prop}.

\head{Application Changes} 
In our design, the application program needs to decide how to store data in three aspects. 
First, the application determines the data that will be used for bulk bitwise operation\fixok{s} so that \fixok{it can inform} the SSD \passthree{to} selectively use \esp for only such data \fixok{(to minimize the storage overhead due to SLC mode)}. 
Second, depending on the computation that can benefit the most from \prop, the application decides whether \passthree{or not} to store the inverse of the original data. 
For example, if the application performs bitwise \bor more frequently than bitwise \band for certain data, it could be more beneficial to store the inverse data to leverage intra-block \mws \passthree{ for bitwise \bor as well}. 
Third, the application decides which operands to be stored in the same block to minimize the number of \mws operations required for the same bitwise operation.

\head{System Software Changes} 
In our design, the application program interacts with the SSD using the Flash-Cosmos library that includes two methods: \inum{i}~\texttt{fc\_write}, which writes the \fixok{operand} data \fixok{for} bitwise operations, and \inum{ii}~\texttt{fc\_read}, which \fixok{reads} the results of bitwise operations. 
Using \texttt{fc\_write}, the application \fixok{informs the SSD of the context of the operation}, such as the programming mode and the location \fixok{(e.g., logical block address)}, to ensure that the data is properly stored for in-flash computation. 
To perform an in-flash bitwise operation, the application uses \texttt{fc\_read} to specify to the SSD the locations of the target operands, the size of the operands, and the types of bitwise operations required.

\head{SSD Changes} In our design, the SSD firmware requires two key changes. 
First, it generates \prop commands to properly handle \texttt{fc\_write} and \texttt{fc\_read} from the host system. 
Second, the SSD firmware maintains additional metadata necessary for \prop, such as each page’s programming mode \fixok{and the location where \passthree{the page should} be stored}.

In this work, \fixok{our focus is} on investigating the feasibility \fixok{and benefits} of \prop \fixok{ using} modern NAND flash memory \fixok{chips}. While the end-to-end support for \prop requires several changes \fixok{within layers of the} system \fixok{stack}, we believe that existing approaches can be applied to meet the key requirements (e.g., efficient storage layout designs~\cite{aga-hpca-2017, seshadri-ieeecal-2015, seshadri-micro-2017, hajinazar-asplos-2021}, host-SSD communication~\cite{hajinazar-asplos-2021, mailthody-micro-2019, kim-asplos-2020}, and metadata management inside the SSD~\cite{mailthody-micro-2019, kim-asplos-2020}). We leave \fixok{more} efficient end-to-end system designs \fixok{and software stack} for Flash-Cosmos to future work.

\section{Methodology}\label{sec:eval_method}

\sloppy\head{Evaluated Systems}
To evaluate the effectiveness of \prop, we \fixok{analyze} four computing platforms:~\inum{i}~an outside-storage processing system (\osp),\inum{ii} an in-storage processing system (\isp), \inum{iii}~\pbit (\pb)~\cite{gao-micro-2021}, and \inum{iv}~\prop (\fc).
\osp performs \bbos using the host CPU concurrently with reading the operands from the SSD to main memory in batches.
\isp leverages an in-storage hardware accelerator that consists of simple bitwise logic and 256-KiB SRAM buffer in order to perform \fixok{\bbos} inside the SSD and send\fixok{s} only the final results to the host.
\pb and \fc perform \bbos inside the NAND flash chips via the in-flash processing mechanisms described in \sect{\ref{ssec:motiv_parabit}} and \sect{\ref{sec:design}}, respectively, and send only the \passthree{final} results to the host.
Unless otherwise specified, we set all the evaluated systems to program (and thus read) the \fixok{inputs and outputs} of bitwise operations in SLC mode for fair performance comparison.

\head{Performance Modeling} 
We use two state-of-the-art simulators to analyze the performance of the evaluated systems. 
We model DRAM timing with \fixok{the} DDR4 interface~\cite{standard-jedec-2020} in Ramulator~\cite{kim-cal-2016, ramulator-github}, a widely-used cycle-accurate DRAM simulator.
We model SSD performance using MQSim~\cite{tavakkol-fast-2018, mqsim-github}, a state-of-the-art SSD simulator. 
We extend MQSim to faithfully model the performance of \isp, \pbit, and \prop with the timing parameters we obtain from our real-device characterization (\sect{\ref{sec:characterization}}). We model the end-to-end throughput of the evaluated systems based on the throughput of each of two \fixok{computation} stages, SSD read (including in-storage processing in \isp, \pb,  and \fc) and host computation (which we measure on a real \passthree{host} system).  
\tab{\ref{tab:config}} \fixok{summarizes} the configurations of the SSD and host system used for our evaluation.

\begin{table}[ht]
    \caption{Evaluated system configurations.}
    \centering
    \resizebox{\columnwidth}{!}{
    \begin{tabular}{@{} c|l @{}}
    \toprule
    \multirow{3}{*}{\textbf{\passthree{Real} Host System}} & \textbf{CPU}: Intel Rocket Lake i7 11700K~\cite{IntelCor20};\\
    & x86~\cite{guide2016intel}; 8 cores; out-of-order; 3.6 GHz;\\\cmidrule{2-2}
    & \textbf{Main Memory}: 64 GB; DDR4-3600; 4 channels;\\
    \midrule
    \multirow{9}{*}{\textbf{\fixok{Simulated SSD}}} & 48-WL-layer 3D TLC NAND flash-based SSD; 2 TB;\\\cmidrule{2-2}
    & \textbf{Bandwidth}: 8-GB/s external I/O bandwidth (4-lane PCIe Gen4);\\
    & 1.2-GB/s Channel IO rate;\\\cmidrule{2-2}
    & \textbf{NAND Config}: 8 channels; 8 dies/channel; 2 planes/dies;\\
    & 2,048 blocks/plane; 196 (4$\times$48) WLs/block; 16 KiB/page;\\\cmidrule{2-2}
    & \textbf{Latencies}: \tr (SLC mode): 22.5 \usec; \tmws: 25 \usec (Max. 4 blocks);\\
    & \tprog (SLC/MLC/TLC mode): 200/500/700 \usec; \tesp: 400 \usec;\\\cmidrule{2-2}
    & \textbf{Power}: HW Accelerator (only in \isp): 93 pJ for 64B operation;\\
   \bottomrule
   \end{tabular}
   }
   \label{tab:config}
\end{table}

\head{Energy Modeling} 
We measure the energy consumption for host computation using Intel RAPL~\cite{hahnel2012measuring}
To model DRAM energy consumption, We use DRAM power values based on \fixok{the} DDR4 model~\cite{ddr4sheet, ghose2019demystifying}.
To model SSD energy consumption, we use the SSD power values of Samsung 980 Pro SSDs~\cite{samsung-980pro} and the NAND flash power values that we measure in our real-device characterization (\sect{\ref{sec:characterization}}).

\head{Workloads}
We evaluate three real-world applications that heavily rely on \bbos.
For fair comparison with \pbit, we evaluate two of the three applications studied in~\cite{gao-micro-2021}, bitmap indices and image segmentation,\footnote{We do not evaluate the other application \passthree{evaluated in \pbit~\cite{gao-micro-2021}}, image encryption~\cite{han-spie-1999}, as it relies \fixok{only} on bitwise \bxor operations that commodity NAND flash chips already support \fixok{(see \sect{\ref{sec:motiv}})}, i.e., neither \pbit nor \prop is necessary to perform in-flash processing for \passthree{such XOR operations}.} as well as a graph-processing workload called k-clique star listing~\cite{besta-micro-2021, danisch-procwwwc-2018, jabbour-kcsl-2018}.
For all workloads, we assume that the data set is initially stored in the SSD due to its large size. 

\noindent
\emph{1) Bitmap Index (\bmi)}: Bitmap indices~\cite{chan-signmod-1998} are an alternative to traditional B-tree indices for databases, which can provide high space efficiency and high performance for many queries (e.g.,~join and scan) compared to B-tree\fixok{s}. 
We assume a database that tracks the log-in activities of $u$ users for a website every day.
For the $i$-th day, the database maintains a vector $D_i$ with $u$ elements, each of which is a 1-bit flag to indicate each user's log-in activity on the day (\texttt{0}: not logged-in, \texttt{1}: logged-in).
Our \bmi workload runs the following query: ``How many users were active every day for the past $m$ months?'' 
Executing the query requires \inum{i}~bitwise \band operations on $d$ vectors, where $d$ is the number of days in the past $m$ months, and \inum{ii}~a {bit-count} operation, i.e., an operation that counts the number of \js{`\texttt{1}' (logged-in)} bits in a given result vector $r$. 
\js{We assume a database that tracks the log-in activities of 800 million users and evaluate the \bmi workload while varying $m$ from 1 to 36.
For executing the \bmi workload, \pb and \fc perform the bit-count operation using the host CPU concurrently with sending the result vector to main memory in batches.}

\noindent
\emph{2) Image Segmentation (\ims)}: 
Image segmentation~\cite{bruce-yuv-2000} is an image processing kernel that aims to break an image into multiple regions depending on a given set of colors.
To determine whether a pixel $p$ belongs to a certain color $C$, our \ims workload uses the YUV color recognition and performs a bitwise \band operation of $Y(p, C)~\bigcdot~U(p, C)~\bigcdot~V(p, C)$, where $Y(p, C)$, $U(p, C)$, and $V(p, C)$ are binary values that can be obtained from pre-processing~\cite{bruce-yuv-2000}.
In our evaluation, \ims segments $I$ images, each of which consists of 800$\times$600 pixels, with four colors. This can be done by performing a bulk bitwise \band operation to three bit vectors where each bit\fixok{-}vector contains $I\times800\times600\times4$ bits. 
We assume the three bit\fixok{-}vectors are initially stored in the SSD and evaluate the \ims workload while varying $I$ from 10,000 to 200,000.

\noindent
\emph{3) $K$-Clique Star Listing (\kcsl)}: $K$-clique star listing~\cite{besta-micro-2021, danisch-procwwwc-2018, jabbour-kcsl-2018} \js{aims} to find all the $k$-clique stars in \fixok{an input} graph.
For a given graph, a $k$-clique is a sub-graph with $k$ vertices that are fully connected to each other.
A $k$-clique star is a collection of \inum{i} a $k$-clique and \inum{ii} all the vertices in the remainder of the graph that are connected to all vertices of the $k$-clique.
Prior work demonstrates that $k$-clique star listing can be significantly accelerated via processing-in-memory with \fixok{a} set-centric formulation~\cite{besta-micro-2021}.
In our evaluation, each vertex is represented using a bit\fixok{-}vector that contains adjacency information to all other vertices in the graph.
Each $k$-clique is represented with another bit vector that specifies the set of vertices that belong to the $k$-clique.
With such bit\fixok{-}vector representations, 
our \kcsl workload can determine a $k$-clique star 
by \fixok{performing} only a bitwise \band operation of the bit\fixok{-}vectors of all the vertices in the corresponding $k$-clique. 
To form the final representation of a $k$-clique star, \kcsl performs a bitwise \bor operation of the calculated intermediate bit vector and the bit\fixok{-}vector that represents the $k$-clique.
Note that \prop can perform both of the bitwise \band and \bor operations simultaneously if the $k$-clique bit vector is stored in a block \fixok{that is} different \fixok{from} than the \fixok{block that stores the} vertex adjacency vectors. We use \fixok{an input graph} with 32 million vertices and 1,024 $k$-cliques and we sweep the dimensions of the cliques\fixok{,} from 8 to 64. 

\fixok{Applications using \bbos with many operands can be particularly sensitive to the RBER of NAND flash memory. For example,  a single bit error in any of the operands in the \bmi workload results in an active user not being counted. The probability of miscounting active users grows as the number of operands increases and rapidly becomes impractical without a sufficiently low RBER.} \fixok{Assuming a best-case RBER of $8.6\times10^{-4}$ (based on our analysis) and $m$=36, the probability of a correct output is 0.42 which is not acceptable. \kcsl is similarly error-intolerant due to the large number of operands \passthree{it uses}. In contrast, \ims is more error tolerant due to the fewer operands in this workload}. \passthree{Thus, we expect that \prop would be a good fit for all these workloads, especially \bmi and \kcsl, due to its zero bit error rate in computation results.}
\section{Evaluation Results}\label{sec:eval_results}
We evaluate the performance and energy efficiency of \prop by comparing its execution time and energy consumption to \fixok{three} baseline computing platforms.

\subsection{Impact on Performance}\label{ssec:eval_perf}

\begin{figure*}[t]
    \centering
    \includegraphics[width=\linewidth]{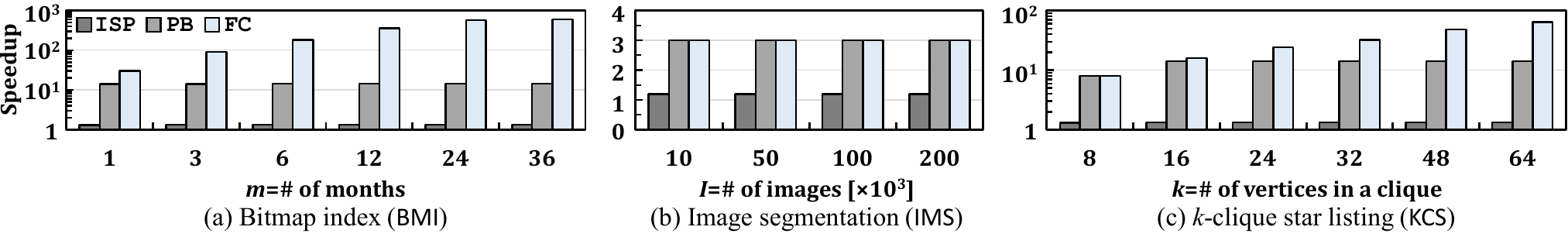}
    \caption{Performance comparison of four computing platforms \passfive{on} three real-world workloads. ISP (In-storage processing), PB (\pbit), FC (\prop). Speedup values are normalized to OSP (Outside-storage processing). Y-axis is in log-scale.}\label{fig:eval_speedup}
\end{figure*} 

\fig{\ref{fig:eval_speedup}} shows the \js{speedups} of \prop~\js{(\fc)},  \pbit~\js{(\pb)}, and the in-storage processing system (\isp) \js{over} the \fixok{conventional} outside-storage processing system \js{(\osp)}. We make six observations from \fig{\ref{fig:eval_speedup}}.

First, \js{\fc significantly} outperforms \osp, providing 32$\times$ speedup on average across all three \js{workloads} and input \js{data} sets.
\js{In \osp, computation can be completely hidden by reading of the operands due to the simple nature of bitwise operations, but the SSD's external bandwidth bottlenecks performance.
This means that, when the operands are stored in \fixok{the} storage system, any other outside-storage processing platform (e.g., GPU or \passthree{near-memory processor}) \emph{cannot} improve the performance of \bbos over \osp \fixok{ (unless one increases SSD's external bandwidth)}.}

Second, \fc also significantly outperforms \isp, providing 25$\times$ speedup on average.
While \isp provides considerable speedup (28\%) over \osp by \fixok{reducing} external data movement \passthree{from the SSD} \fixok{(external bandwidth is 8 GB/s, \tab{\ref{tab:config}})}, the limited \emph{internal} \fixok{SSD} bandwidth \fixok{(9.6 GB/s, \tab{\ref{tab:config}})} becomes a new bottleneck.
\isp needs to read out \emph{all} operands from NAND flash chips to the hardware accelerators, which can be largely avoided by in-flash processing.

Third, \fc provides large performance improvement \fixok{over} \pbit.
While \pb also significantly outperforms \osp{} (and \isp) by 9.4$\times$ (7.2$\times$), \fc \fixok{ outperforms} \pb by 3.5$\times$ on average across all three \js{workloads} and input \js{data} sets.
This highlights the key benefit of performing \mws in real-world applications. 

Fourth, the benefits of \fc increase with the number of operands \passthree{used in} \bbos.
The speedups of \fc for \bmi are higher compared to the other two workloads, since \bmi has a larger number of operands (from 30 to 1,095).
\passfour{In contrast, the performance of \pb does not improve as the number of operands increases (e.g., for $k$$>$16 in \kcsl).}
This is because the performance bottleneck \passfour{of \pb} shifts to \passfour{the serial sensing} of the operands as discussed in \sect{\ref{ssec:motive_limit}}, due to which the latency also linearly increases \fixok{with} the number of operands.  
As a result, while \fc significantly improves performance over \osp/\isp for \bmi by 198.4$\times$/150.5$\times$, \pb's benefits over \osp/\isp for \bmi remain \passthree{at only} 14$\times$/$10.7\times$.

Fifth, the benefits of \fc are affected by not only the number of operands but also the operand size. \passfour{For example, \fc does the same  operation (\texttt{AND}) over 30 operands for \bmi (when $m$=1) and  over 32 operands for \kcsl (when $k$=32). Although \kcsl has more operands than \bmi, we observe that \fc \passfive{ provides} higher improvement over \pb for \bmi.} %
This is because the total size of \passfour{the} result bit vectors is smaller in \bmi (100 MB) than in \kcsl (4 GB)\passthree{, which results in reduced \passfour{external I/O time spent for transferring the results from the SSD to the host}}.
\passthree{As shown in \fig{\ref{fig:motiv_ifp}(d)}, external data movement can become the bottleneck even for in-flash processing if the external I/O time \passfour{within an application (e.g., due to transferring results out of the SSD) is} larger than the overall sensing time. \passfour{Therefore, the benefit of \fc over \pb is lower if the external I/O time dominates the overall execution time.}}

Sixth, \fc and \pb show almost similar performance for \ims across all input sets. 
Even though \fc reduces the average number of sensing operations by 3$\times$ \passthree{compared to \pb} when executing the \ims workload, moving the large (up to 44GiB) result vector \passthree{of \ims} to the host dominates the total execution time for both mechanisms. This is in contrast to \bmi where the vector size is only 100 MB.
\passfour{Note that, in \ims, both \fc and \pb} provide high speedups compared to \isp and \osp (2.5$\times$ and 3$\times$, respectively) by reducing the amount of external and internal data transfers, which \passthree{are} the performance bottleneck in \ims.

We conclude that \prop is an effective substrate to accelerate important real-world applications. \passthree{\prop not only largely} outperforms the state-of-the-art IFP technique \passthree{but also, crucially,} provides reliable, \passthree{error-free} execution (which is necessary for correct results in all three real-world workloads we evaluate).

\subsection{Impact on Energy Consumption}\label{ssec:eval_energy}
\begin{figure*}[h]
    \centering
    \includegraphics[width=\linewidth]{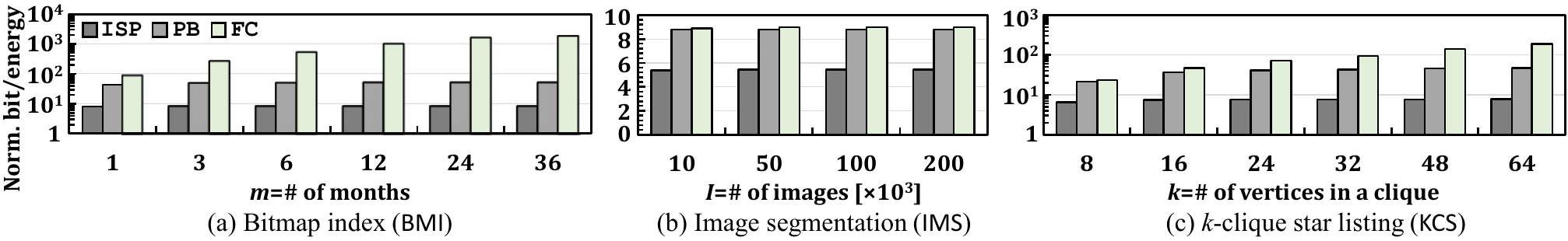}
    \caption{Energy-efficiency comparison of four computing platforms \passfive{on} three real-world workloads. Normalized to OSP (Outside-storage processing). Y-axis represents number of bits computed per unit of energy in log-scale.}\label{fig:eval_energy}
\end{figure*}

Fig.~\ref{fig:eval_energy} shows the energy-efficiency \fixok{of} \fc, \pb, and \isp\fixok{,} \fixok{in terms of \passthree{the} number of bits that can be computed/transferred per unit of energy,} normalized to that of \osp. We make three observations. First, \fc greatly increases energy efficiency over the other evaluated systems, providing 95$\times$/13.4$\times$/3.3$\times$ \passthree{higher} \passthree{energy-efficiency} compared to \osp/\isp/\pb, on average across all three \fixok{workloads} and input sets. \prop has the maximum energy savings of 1,839$\times$/222$\times$/35.5$\times$ over \osp/\isp/\pb for \bmi when $m$=36.
Second, the overall trend of \fc's energy-efficiency benefits is similar to its performance benefits. The \passthree{energy} benefits of \fc increase (decrease) as the number of operands (the operand size) increases.
\passthree{Third, \fc reduces energy by not only reducing data movement but also reducing sensing energy, especially for \passfour{multi-operand} operations. As a result, it provides higher improvements in energy efficiency (95$\times$ on average over \osp) \fixok{than} in performance (32$\times$ on average over \osp)}.
Note that, as shown in \figs{\ref{fig:eval_speedup}(b) and \ref{fig:eval_energy}(b)}, \passthree{for \ims, even though} \fc \passthree{ performs similarly to \pb, it provides 2.3\% energy savings}.
We conclude that \prop is an efficient substrate to eliminate \fixok{the energy overheads of} data movement for many commonly-used real-world applications.

\subsection{Overhead Analysis}\label{ssec:overhead}
As briefly discussed in \sect{\ref{ssec:idea_esp}}, \prop introduces two key overheads due to the use of \esp for reliable in-flash computation.
First, \esp requires 2$\times$ \fixok{the} page-program latency compared to regular SLC-mode programming. 
Second, \esp consumes 2$\times$ \fixok{the storage} capacity to store the same amount of data compared to MLC-mode programming.

While the write-performance and capacity overheads are not negligible, we believe that both overheads would not be serious obstacles to the adoption of \prop due to three reasons. 
First, \esp is essential to ensuring the reliability of in-flash computation.\footnote{\passthree{In order to} be \passthree{useful} for general-purpose computation, \pbit has to support \passthree{error-free computation}, potentially using \esp.}
As shown in \sect{\ref{ssec:motive_limit}}, regular SLC/MLC-mode programming exhibits significantly higher RBER than the UBER requirement\passthree{, and \esp \passfour{ effectively} solves this \passfour{important} reliability problem \passfour{that is present in} in-flash computation}.\footnote{\prop can also work with MLC NAND flash memory while guaranteeing the same level of reliability as \pbit provides, when the operands are stored in LSB pages. 
This is because the mechanism of LSB-page reads is the same as SLC-page reads, except for the used read-reference voltage levels (LSB-page read in MLC: \vrefn{2} in in \fig{\ref{fig:bg_vth}(b)} vs. SLC-page read: \vref in \fig{\ref{fig:bg_vth}}(a)).}

Second, both \fixok{the write-performance and capacity} overheads \fixok{apply \emph{only}} to the data used for bulk bitwise operations.
\fixok{As such, \prop can minimize these overheads by selectively using \esp only for the data that is involved in in-flash processing (\bbos) while programming other data using regular SLC/MLC/TLC-mode programming. Such a functionality is supported by the multiple programming modes in modern NAND flash memory, i.e., any block can be programmed in SLC, MLC, and TLC modes~\cite{lee-atc-2009, kim-asplos-2020, li-iccad-2019}.} 

Third, \esp \passfour{ does} not degrade SSD write performance, in terms of both bandwidth and latency, compared to MLC-mode programming. 
This is because the program latency of \esp (400~\usec, \passthree{\sect{\ref{ssec:idea_esp}}}) is still lower than \fixok{the latency} of MLC- and TLC-mode programming (500~\usec and 700~\usec, respectively, in our evaluated chips).
We evaluate the sequential write bandwidth of \esp, and the results show that \esp provides a write bandwidth of 4.7 GB/s, which is 73.4\%/121.4\%/166.7\% \fixok{of the} regular SLC/MLC/TLC-mode programming \fixok{write bandwidth} (6.4/3.87/2.82 GB/s).

\section{Related Work}\label{sec:related}
 To our knowledge, this work is the first to enable in-flash bulk bitwise operations on \emph{multiple operands} through a \emph{single sensing} operation, while achieving \emph{zero bit errors} in the computation results. 
 \fixok{We} already discussed \fixok{and comprehensively compared to} the state-of-the-art technique\passsix{~\cite{gao-micro-2021}} closely related to \prop \passsix{ (Sections {\ref{sec:motiv}}, {\ref{sec:eval_method}}  and \ref{sec:eval_results})}. We briefly \fixok{describe other} NDP proposals at different levels \fixok{in} the memory hierarchy.

\head{In-Flash Processing} 
\fixok{Several} prior works propose in-flash processing techniques to accelerate \fixok{the multiply-and-accumulate (MAC) operations in } different applications such as neural networks \passfive{(e.g.,\cite{choi-iscas-2020, han-ieeetcas-2019, lue-iedm-2019, wang-ieeetvlsi-2018, wang-edtm-2022,han-aisy-2021,kang-tc-2021, lee-fnins-2020,kim-fast-2021})} and mixed signal sensing (e.g.,\cite{bayat-ieeetnnls-2017}). Similar to \prop, these mechanisms have high bit-level parallelism, but they exploit analog current accumulation, \passthree{which} requires significant changes to the NAND flash cell array structures, e.g., deploying precise \passthree{and costly} analog-to-digital converters inside the chip (see \sect{\ref{ssec:idea_mws})}. In contrast, our mechanism can be adopted in commodity SSDs with very low cost, as we demonstrated in \sect{\ref{sec:characterization}}.

\head{In-Storage Processing} Several prior works propose to leverage the internal processor \passfive{(e.g.,~\cite{seshadri-osdi-2014,mansouri-asplos-2022,kang-msst-2013,gu-isca-2016,wang-eurosys-2019,acharya-asplos-1998,keeton-sigmod-1998, wang-damon-2016, koo-micro-2017, tiwari-fast-2013, tiwari-hotpower-2012, boboila-msst-2012, bae-cikm-2013,torabzadehkashi-ipdpsw-2018, kang-micro-2021,li-atc-2021,lim-icce-2021,kim-sigops-2020})} or embed hardware accelerators \passfive{(e.g.,~\cite{jun-isca-2018, mailthody-micro-2019,pei-tos-2019, do-sigmod-2013,kim-infosci-2016, riedel-computer-2001,riedel-vldb-1998, torabzadehkashi-pdp-2019,liang-atc-2019,cho-wondp-2013, jun-isca-2015, lee-ieeecal-2020, ajdari-hpca-2019,liang-fpl-2019,jeong-tpds-2019,jun-hpec-2016})} within the storage device for computation. Due to \fixok{their} more \fixok{general-purpose} designs, these proposals can perform more diverse \fixok{and complex} operations (e.g., arithmetic operations). However, as discussed in \sect{\ref{sec:motiv}}, in-storage processing needs to first read out the processed data from the \fixok{flash chips} and transmit it to the SSD controller over the \fixok{SSD-}internal I/O link\fixok{, which is a performance bottleneck.} In contrast, \prop can effectively \fixok{reduce the data movement between NAND flash chips and the SSD controller \passthree{by performing computation inside the flash chip arrays, leading to more than an order of magnitude higher performance and energy-efficiency than in-storage processing} (\sect{\ref{sec:eval_results}}).} 

\head{In-Memory and In-Cache Processing} A large body of prior work proposes various NDP techniques \fixok{at other} levels of the memory hierarchy, e.g., \passfive{in main memory (e.g.,~\cite{seshadri-micro-2017, seshadri-ieeecal-2015, li-micro-2017, li-dac-2016, angizi-glsvlsi-2019,mutlu-emergingcomputing-2021,ali-tcas-2019,chang-hpca-2016,deng-dac-2018,gao-micro-2019,seshadri-micro-2013,xin-hpca-2020,seshadri-arxiv-2016,seshadri-arxiv-2019, hajinazar-asplos-2021,truong-micro-2021,giannoula-pomacs-2022,dai-tcad-2019,tang-aspdac-2017,leitersdorf-arxiv-2022,rohbani-dac-2022, perach-arxiv-2022,li-jcst-2021,kim-hpca-2021,zhang-hpdc-2014,santos-date-2017, ferreira-arxiv-2021,zhuo-micro-2019, patterson-ieeemicro-1997, boroumand-icde-2022, singh-dac-2019,seshadri-arxiv-2016-pum, boroumand-ieeecal-2017,ping-isca-2016,kautz-tc-1969,stone-tc-1970})} and in \passfive{SRAM caches (e.g.,~\cite{aga-hpca-2017, fujiki-isca-2019, kang-jproc-2020, angstadt-micro-2018,eckert-isca-2018, kang-procieee-2020,zhang-jssc-2017,lin-vlsidat-2021,alhawaj-iscas-2020,zhang-arxiv-2022})}. \passthree{Even though these works provide} significantly lower access latency and high reliability, once the size of the processed data exceeds the cache \passthree{and main memory} capacity, the data needs to move between the storage \passthree{devices} and the rest of the memory hierarchy. \prop can complement these NDP approaches (including in-storage processing) by processing \fixok{large amounts of} data \fixok{inside flash arrays} and \fixok{communicating} only the results of the computation.

\section{Discussion}\label{sec:discussion}
\head{Extensions \fixok{to} Other Applications}
\prop can be used to accelerate not only bitwise operations but also any desired operation. 
This is because \prop supports a set of bitwise operations that are logically complete, like other \emph{processing-using-memory (PuM)} substrates that use the operational principles of the memory cells for computation, such as Compute Caches~\cite{aga-hpca-2017} (SRAM-based PuM), Ambit~\cite{seshadri-micro-2017} (DRAM-based PuM), and Pinatubo~\cite{li-dac-2016} (NVM-based PuM). 
Follow-up works (e.g., DualityCache~\cite{fujiki-isca-2019}, SIMDRAM~\cite{hajinazar-asplos-2021}, and IMP~\cite{fujiki-asplos-2018}) propose frameworks that leverage \fixok{these} substrates and techniques to automate the creation of desired complex operations (e.g., addition and multiplication) to accelerate a broad range of workloads, including graph processing, databases, \fixok{neural networks} and \passthree{genome analysis}. 
We leave the development of such a framework for \prop to \hypertarget{SC8c}{future work}.

\head{Limitations}
\prop has two key limitations that \passthree{also} commonly exist in other PuM solutions. 
First, like \pbit and other PuM proposals \passfive{(e.g., ~\cite{aga-hpca-2017, seshadri-micro-2017, li-dac-2016, seshadri-ieeecal-2015, seshadri-micro-2013, seshadri-arxiv-2019,seshadri-arxiv-2016, hajinazar-asplos-2021,angizi-glsvlsi-2019, truong-micro-2021,ali-tcas-2019,chang-hpca-2016,deng-dac-2018,gao-micro-2019,li-micro-2017,seshadri-arxiv-2016-pum})}, it is not straightforward \fixok{for \prop} to work with mainstream encryption techniques (e.g., AES-256~\cite{biryukov-asiacrypt-2009, samsung-980pro}) \fixok{that are} widely used in modern SSDs. 
This is because widely-used encryption techniques have input-data dependence and/or require complex computation other than bitwise operation\fixok{s} (e.g., shift\fixok{ing}). 
One possible solution is to employ homomorphic encryption that preserves the correctness of computation for encrypted data~\cite{fontaine-eurasipjis-2007}. 
Although homomorphic encryption currently has many challenges \passthree{with large} computation and capacity overheads, we believe that the development of efficient homomorphic encryption would be a promising direction to solve \fixok{this} common problem of \fixok{the} PuM \hypertarget{ICB1c}{paradigm} \fixok{in dealing with encrypted data}.

Second, like \pbit and other PuM proposals, \prop can accelerate bulk bitwise operations only when the operands are stored in the same chip. 
The system can potentially leverage an efficient inter-chip data migration technique to gather the target operands \fixok{into} the same block in background, but doing so inevitably incurs data movement that \fixok{eats away from} the benefits of Flash-Cosmos. 
When the operands are stored in different chips, in-storage processing that uses hardware accelerators near NAND flash chips could be more effective. 
Fortunately, \prop requires only small changes to commodity NAND flash chips, which makes it easy to be combined with such \fixok{an} in-storage processing solution. 
We leave the integration of Flash-Cosmos with other NDP solutions to future work. 

\section{Conclusion}\label{conclusion}
We propose \prop, a new \fixok{in-flash processing} technique that significantly improves the performance, energy efficiency, and reliability of in-flash bulk bitwise operations.
\prop takes full advantage of the massive bit-level parallelism present in modern NAND flash memory by leveraging the cell-array structures and operating principles of NAND flash memory. \passthree{First, \prop} enables the chips to perform \bbos on multiple (tens) operands via \passthree{only one} single-sensing \passthree{operation}. \passthree{Second,} \prop enhances \passthree{the} existing SLC-mode programming scheme to achieve \fixok{zero bit errors in computation results}, thereby \fixok{enabling the use of} in-flash processing \fixok{for} general\fixok{,} error-intolerant applications\fixok{, which was previously not possible}.
We experimentally demonstrate the feasibility, performance, and reliability of \prop using 160 real 3D NAND flash chips. \passfour{Our} simulation-based \passthree{real workload} evaluations show that \prop significantly outperforms \passthree{outside-storage processing, in-storage processing and the state-of-the-art in-flash processing \passfour{technique}} in terms of both performance and energy efficiency while providing reliable operation. We conclude that \prop is a promising substrate to enable highly-efficient, high-performance, \passthree{and reliable} in-flash computation. We hope and expect that future work builds on \prop in many ways, e.g., by enabling system-level frameworks that take advantage of \prop and by demonstrating benefits over more workloads.

\section*{Acknowledgments}
We thank the anonymous reviewers of ISCA 2022 and MICRO 2022 for feedback. 
We thank the SAFARI group members for feedback and the stimulating intellectual environment. 
We specifically thank Nika Mansouri Ghiasi and Minesh Patel who helped shape our arguments and strengthen our evaluation.
We acknowledge the generous gifts and support provided by our industrial partners: Google, Huawei, Intel, Microsoft, VMware, the Semiconductor Research Corporation and the ETH Future Computing Laboratory. 
Jisung Park was in part supported by the National Research Foundation (NRF) of Korea (NRF-2020R1A6A3A03040573).
\emph{(Co-corresponding Authors: Jisung Park, Myungsuk Kim, and Onur Mutlu)}

\balance

\bibliographystyle{IEEEtran}

\newpage

\end{document}